\documentclass[preprint]{revtex4-1}

\usepackage{dcolumn}
\usepackage{bm}
\usepackage{amssymb}
\usepackage{subfigure}
\usepackage{here}
\usepackage{ulem}
\usepackage{enumerate}
\usepackage{cancel}
\usepackage{flushend}
\usepackage{bbold}
\usepackage{mathtools}
\usepackage{float}
\usepackage{stmaryrd}
\usepackage{colortbl}
\usepackage[table]{xcolor}
\usepackage{array,ragged2e}
\usepackage{amsmath,amssymb,amsfonts,stmaryrd,wasysym,graphicx,multirow,color,textcomp,subfigure}
\usepackage{url}
\usepackage[colorlinks=true,citecolor=blue,urlcolor=blue]{hyperref}
\usepackage{romannum}
\usepackage{nicematrix}
\usepackage{tabularx}
\usepackage{changes}

\usepackage[utf8]{inputenc}
\usepackage{tikz-cd}

\usepackage{graphicx}
\usepackage{color}

\newcommand{\deri}{\text{d}}
\newcommand{\imag}[1]{\text{Im}($#1$)}

\newcommand{\eq}[1]{Eq.~(\ref{#1})}

\newcommand{\tbf}[1]{\textbf{#1}}

\newcommand*{\thead}[1]{\multicolumn{1}{c}{\bfseries #1}}

\newenvironment{helv}{\fontfamily{phv}\selectfont}{\par}
\DeclareTextFontCommand{\texthelv}{\helv}

\makeatletter
\newcommand{\thickhline}{%
    \noalign {\ifnum 0=`}\fi \hrule height 1pt
    \futurelet \reserved@a \@xhline
}
\newcolumntype{"}{@{\hskip\tabcolsep\vrule width 1pt\hskip\tabcolsep}}
\makeatother

\begin{document}

\draft
\title{Classification of multiple arbitrary-order non-Hermitian singularities}
\author{Jung-Wan Ryu}
\address{Center for Theoretical Physics of Complex Systems, Institute for Basic Science (IBS), Daejeon 34126, Republic of Korea.}
\author{Jae-Ho Han}
\address{Center for Theoretical Physics of Complex Systems, Institute for Basic Science (IBS), Daejeon 34126, Republic of Korea.}
\author{Chang-Hwan Yi}
\email{yichanghwan@hanmail.net}
\address{Center for Theoretical Physics of Complex Systems, Institute for Basic Science (IBS), Daejeon 34126, Republic of Korea.}
\date{\today}

\begin{abstract}
 We demonstrate general classifications of Riemann surface topology generated by multiple arbitrary-order exceptional points of quasi-stationary states. Our studies reveal all possible product permutations of holonomy matrices that describe a stroboscopic encircling of 2nd order exceptional points. The permutations turn out to be categorized into a finite number of classes according to the topological structures of the Riemann surfaces. We further show that the permutation classes can be derived from combinations of cyclic building blocks associated with higher-order exceptional points. Our results are verified by an effective non-Hermitian Hamiltonian founded on generic Jordan forms and then examined in physical systems of desymmetrized optical microcavities.
\end{abstract}

\maketitle

\section{Introduction}

{\it Exceptional points -} Non-Hermiticity is ubiquitous in many branches of physics such as quantum mechanics, optics, condensed matter physics, and nonlinear dynamics~\cite{Moi11, Gan18, Ash20, Ber21}. Known to exhibit many bizarre phenomena compared to Hermitian systems, non-Hermitian physics includes fascinating topics that are deeply related to a non-Hermitian degeneracy, the so-called exceptional point (EP)~\cite{Kat66, Hei90, Hei04}, where eigenvalues and corresponding eigenvectors coalesce simultaneously in parameter space. Because of this defectiveness, singularities at EPs induce non-trivial physical observables~\cite{Lee08} and abnormal topological structures~\cite{Dem01}. Particularly, the topological structures of a Riemann surface around EPs result in a non-orientable behavior for a stroboscopic encircling of EPs: one state changes into another state after one loop around 2nd order EPs, and then recovers its initial state after circling once more. Thus far, the physical advantages of EPs have been utilized in many applications, e.g., orbital angular momentum microlasers~ \cite{Mia16}, ferromagnetic bilayers~\cite{Yan18}, non-Hermitian lattices \cite{Yan21}, and diverse systems in optics and photonics~\cite{Che17, Mir19}.

{\it Multiple exceptional points -} Topological structures of the Riemann surfaces around multiple EPs persist unless we add EPs or remove the initially embedded EPs. The structures can be characterized by factorization sets of holonomy matrices that describe the stroboscopic encircling of each different EP~\cite{Car09, Ryu12}. Here, the topological notion of homotopy with fixed endpoints explains the equivalence among encircled loops~\cite{Zho18}. To date, there have been many studies on the encircling of a few EPs in single-particle systems~\cite{Din15, Din16, Cui19, Jia20, Tan20} as well as on that of a large number of EPs in many-body systems~\cite{Lui19}.

{\it Higher-order exceptional points -} Higher-order EPs, at which many eigenvalues (and eigenvectors) coalesce, have been intensively studied due to their extraordinary characteristics that are highly fruitful in many applications, e.g., sensors~\cite{Hei08, Dem12, Hei16, Sch17}. This is because the Riemann surfaces around $N$th order EPs generally follow multi-valued functions of $f(z) = z^{1/N}$. Note, however, that there is also an exotic class of 3rd order EPs~\cite{Dem12}. So far, higher-order EPs have been achieved in various systems including ring resonators~\cite{Hod17}, coupled-resonator optical waveguides~\cite{Nad17}, weakly deformed microdisks~\cite{Kul18, Kul19}, electronic circuits~\cite{Xia19}, unidirectionally coupled photonic spin-orbit systems~\cite{Wan19}, supersymmetric arrays~\cite{Zha20}, non-Hermitian lattices~\cite{Xia20}, and ferromagnetic trilayers~\cite{Yu20}. Additionally, a systematic methodology to achieve higher-order EPs in low-dimensional parameter space has been proposed~\cite{Qui19, Zho20, Wan20}, and higher-order EPs have been studied in terms of non-Hermitian skin effects as well~\cite{Alv18}.

\begin{figure*}[t]
\begin{center}
\includegraphics[width=.8\textwidth]{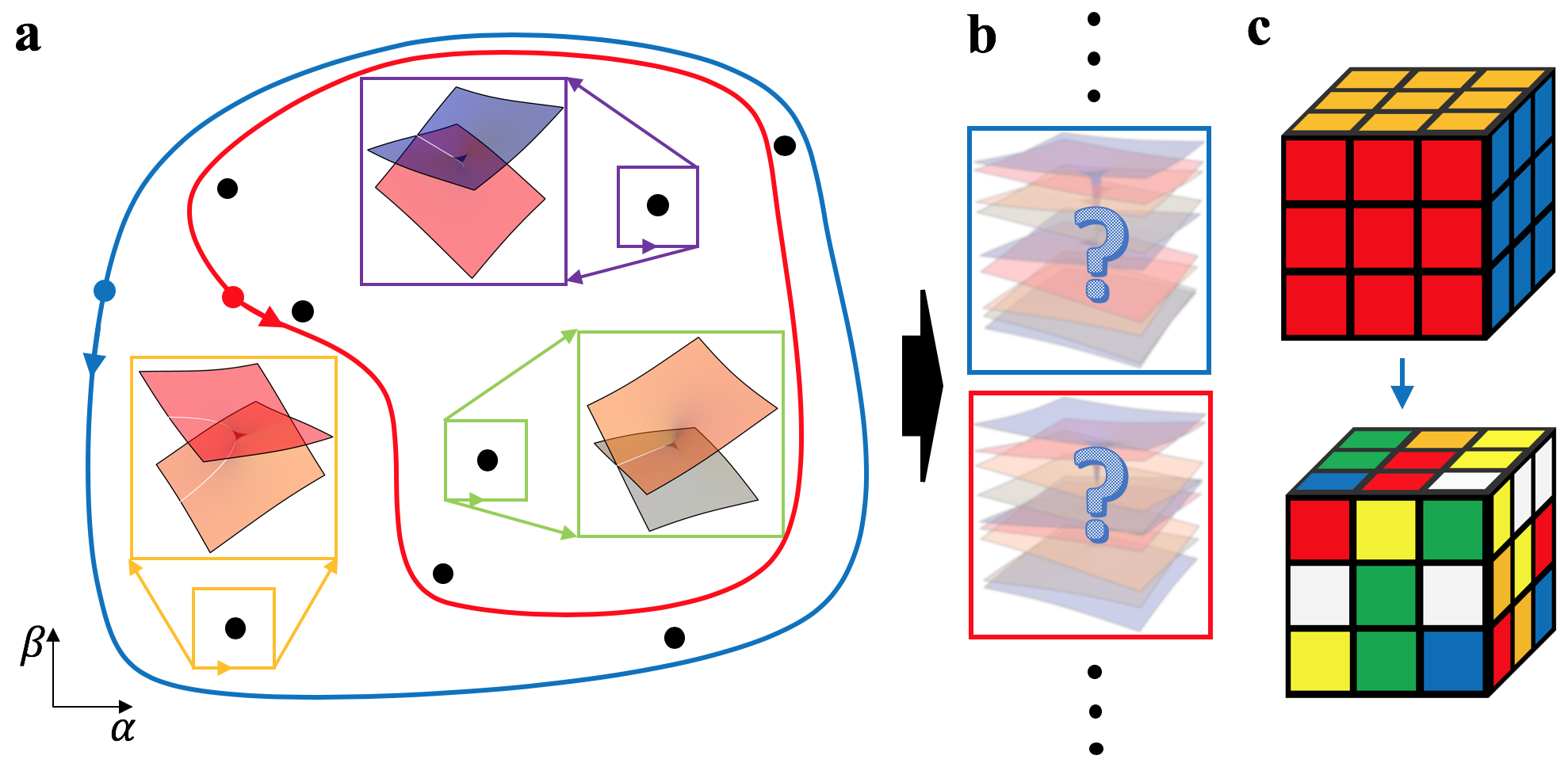}
\caption{{\bf a.} Three selected Riemann surfaces showing state exchanges by closed loops encircling single 2nd order EPs (orange, green, and violet boxes) in a two-dimensional parameter space ($\alpha$,$\beta$). An individual 2nd order EP has its own Riemann surface connecting two corresponding states. The blue and red closed loops represent two loops encircling multiple 2nd order EPs. Parameter sets change adiabatically along the loops from the initial sets (blue and red dots) and then finally return to their initial sets. {\bf b.} Classification of the Riemann surfaces encircling arbitrary numbers of 2nd order EPs associated with $N$ states. {\bf c.} The states are shuffled after encircling multiple 2nd order EPs, operations that form a permutation group analogous to rotations of a Rubik's Cube.}

\label{question}
\end{center}
\end{figure*}

Despite the numerous preceding works on multiple 2nd order EPs or higher-order EPs under specific constraints, a thorough investigation into the Riemann surface classifications of multiple arbitrary-order EPs has yet to be conducted (see Fig.~\ref{question}). With this motivation, we study the topological structures of the Riemann surfaces around multiple numbers of arbitrary-order EPs in terms of the permutation properties of the stroboscopic encircling of EPs. The encircling of EPs can be characterized by the so-called holonomy matrices associated with the permutation groups~\cite{Ham62}. We analyze all possible classes of topological structures of Riemann surfaces around one or multiple arbitrary-order EPs in terms of the holonomy matrix products that describe the 2nd order EP permutations and the combinations of $N$th order EPs of disjoint cycles. Here, two essential properties of the permutation groups are postulated: (i) {\it every permutation can be written as a product of transpositions}, and (ii) {\it any permutation can be expressed as the product of disjoint cycles}.

\section{Results}

\subsection{Permutation as a product of transpositions}

{\it 2nd order EP -} While 2nd order EPs are easy to find as they are persistent in two-dimensional parameter space, regardless of the system symmetries, obtaining higher-order EPs is in general an extremely demanding task due to the need for fine-tuning many parameters. To obtain an $N$th order EP, one needs $(N^2+N-2)/2$ parameters to be tuned~\cite{wdheiss}, e.g., five parameters for a 3rd order EP. When the system has certain symmetries, however, the number of tuning parameters is reduced, and higher-order EPs can appear in lower-dimensional parameter spaces. For instance, in parity and time-reversal (PT) symmetric systems \cite{Ben98}, a 2nd order EP can be formed by a single parameter control.

If these kinds of symmetries are spoiled, the higher-order EPs split into multiple lower-order EPs. Moreover, in perfectly desymmetrized systems, the higher-order EPs disappear except for accidental cases, and in principle, only 2nd order EPs can survive in two-dimensional parameter space. Yet, if the desymmetrizing disorders are not so strong, these 2nd order EPs are still in the vicinity of the original positions of the higher-order EPs in the parameter space. Around such a position, the eigenvalue structures can be complicated. In contrast, those far from this area are topologically protected and well defined as like the Riemann surfaces of the original higher-order EPs. Henceforth, we abbreviate $N$th order EPs as EP$N$s.

\begin{figure*}
\begin{center}
\includegraphics[width=0.7\textwidth]{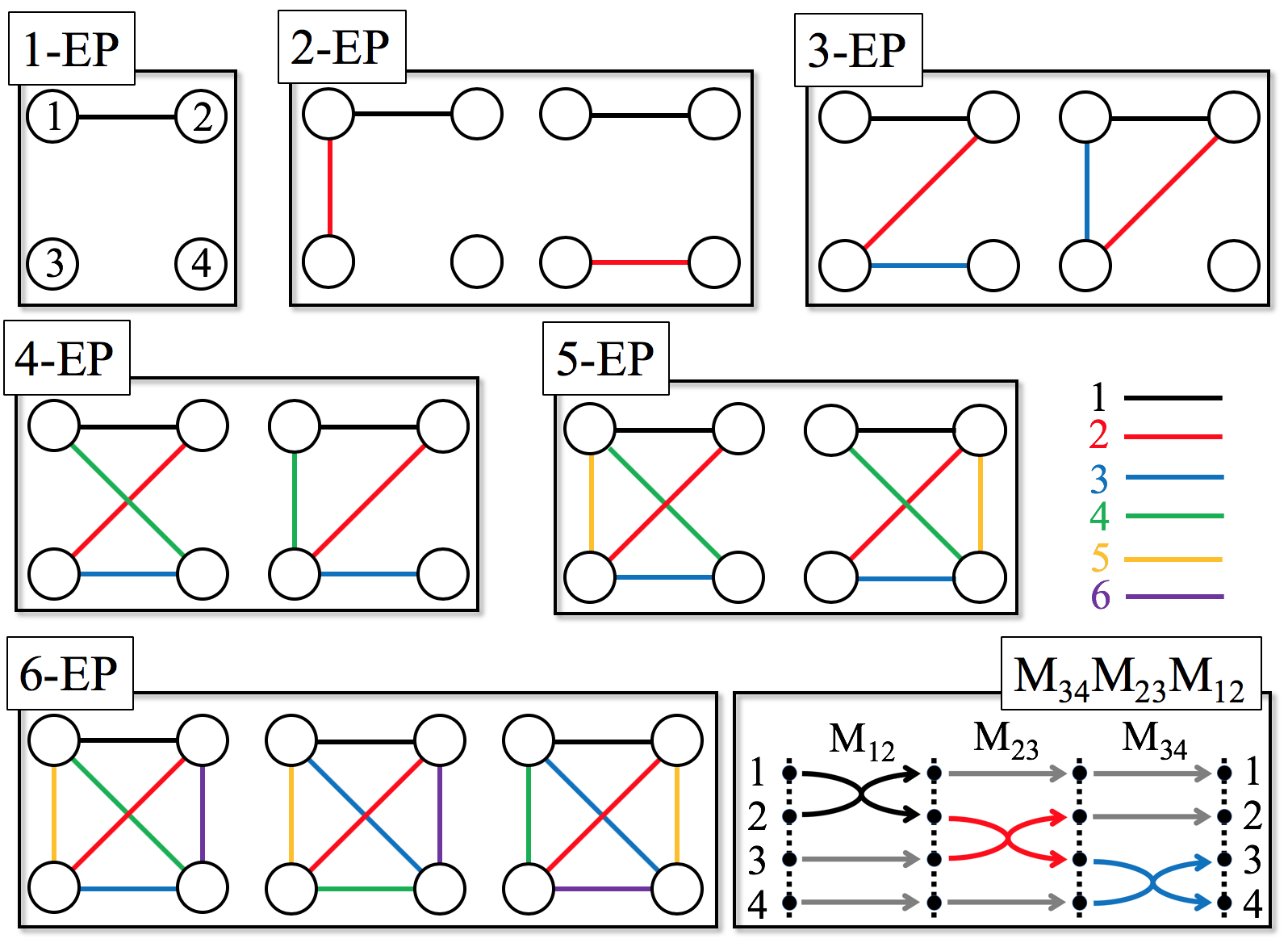}
\caption{Diagram of four states (circles) and six associated 2nd order EPs (colored lines). The line colors represent the permutation order of the holonomy matrices as follows: black, red, blue, green, orange, and violet represent the 1st, 2nd, 3rd, 4th, 5th, and 6th holonomy matrices. The permutations are selected examples corresponding to those in Table~\ref{table_4}. (1-EP) $\mathrm{\bf M}_{12}$. (2-EP) $\mathrm{\bf M}_{13}\mathrm{\bf M}_{12}$ and $\mathrm{\bf M}_{34}\mathrm{\bf M}_{12}$. (3-EP) $\mathrm{\bf M}_{34}\mathrm{\bf M}_{23}\mathrm{\bf M}_{12}$ and $\mathrm{\bf M}_{13}\mathrm{\bf M}_{23}\mathrm{\bf M}_{12}$. (4-EP) $\mathrm{\bf M}_{14}\mathrm{\bf M}_{34}\mathrm{\bf M}_{23}\mathrm{\bf M}_{12}$ and $\mathrm{\bf M}_{13}\mathrm{\bf M}_{34}\mathrm{\bf M}_{23}\mathrm{\bf M}_{12}$. (5-EP) $\mathrm{\bf M}_{13}\mathrm{\bf M}_{14}\mathrm{\bf M}_{34}\mathrm{\bf M}_{23}\mathrm{\bf M}_{12}$ and $\mathrm{\bf M}_{24}\mathrm{\bf M}_{14}\mathrm{\bf M}_{34}\mathrm{\bf M}_{23}\mathrm{\bf M}_{12}$. (6-EP) $\mathrm{\bf M}_{24}\mathrm{\bf M}_{13}\mathrm{\bf M}_{14}\mathrm{\bf M}_{34}\mathrm{\bf M}_{23}\mathrm{\bf M}_{12}$, $\mathrm{\bf M}_{24}\mathrm{\bf M}_{13}\mathrm{\bf M}_{34}\mathrm{\bf M}_{14}\mathrm{\bf M}_{23}\mathrm{\bf M}_{12}$, and $\mathrm{\bf M}_{34}\mathrm{\bf M}_{24}\mathrm{\bf M}_{13}\mathrm{\bf M}_{14}\mathrm{\bf M}_{23}\mathrm{\bf M}_{12}$. ($\mathrm{\bf M}_{34}\mathrm{\bf M}_{23}\mathrm{\bf M}_{12}$) Example of permutations of holonomy matrices for three EPs in a four-state case.}
\label{meps}
\end{center}
\end{figure*}

{\it Stroboscopic encircling of EPs as a permutation group -} The stroboscopic parametric encircling of EPs is a bijection between sets of initial and outcome eigenstates. Considering a system described by an effective $N\times N$ Hamiltonian, off the EPs, there are $N$ eigenstates, and the states are labeled $\{1, 2, \cdots, N\}$ in a specific order. Then a bijection of the parametric encircling of the EPs forms a permutation group of degree $N$, $S_N$. For example, the eigenstate $\{1,2\}$ of a $2 \times 2$ effective Hamiltonian changes into the state $\{2,1\}$ after encircling EP2 once. Together with the identity, which is obtained by encircling EP2 twice, the group $S_2$ is formed. The switch can be described by the holonomy matrix
\begin{eqnarray}
{\mathrm{\bf M}_{12}} = 
\begin{pmatrix} 
 0 & 1 \\
 1 & 0 \\
\end{pmatrix}\ ,
\end{eqnarray}
which is a unitary matrix connecting two eigenstates, $(1,0)^T$ and $(0,1)^T$. While the unsigned holonomy matrix shown above describes the topological structures only for the eigenvalues around EP2s, a signed one can account for the geometric phase of the eigenstates as well (see Supplementary Information). Note that signed holonomy matrix permutations have braiding directions depending on the signs. In this work, we only consider unsigned holonomy matrices.

To be concrete, this work mainly focuses on multiple EP2s associated with EP4 in a $4 \times 4$ effective Hamiltonian (see Supplementary Information for more details). This system is simple but yields spectral structures rich enough to clearly reveal our main results. We then extend our classification formula to general $N$th order cases. The holonomy matrices switching two of four eigenstates after one parametric encircling of EP2 can be given as
\begin{eqnarray}
{\mathrm{\bf M}_{12}} = 
\begin{pmatrix} 
 0 & 1 & 0 & 0 \\
 1 & 0 & 0 & 0 \\
 0 & 0 & 1 & 0 \\
 0 & 0 & 0 & 1 \\
 \end{pmatrix},\ 
 {\mathrm{\bf M}_{23}} = 
\begin{pmatrix} 
 1 & 0 & 0 & 0 \\
 0 & 0 & 1 & 0 \\
 0 & 1 & 0 & 0 \\
 0 & 0 & 0 & 1 \\
 \end{pmatrix},\
 {\mathrm{\bf M}_{34}} = 
\begin{pmatrix} 
 1 & 0 & 0 & 0 \\
 0 & 1 & 0 & 0 \\
 0 & 0 & 0 & 1 \\
 0 & 0 & 1 & 0 \\
 \end{pmatrix},\\\nonumber
 {\mathrm{\bf M}_{14}} = 
\begin{pmatrix} 
 0 & 0 & 0 & 1 \\
 0 & 1 & 0 & 0 \\
 0 & 0 & 1 & 0 \\
 1 & 0 & 0 & 0 \\
 \end{pmatrix},\
 {\mathrm{\bf M}_{13}} = 
\begin{pmatrix} 
 0 & 0 & 1 & 0 \\
 0 & 1 & 0 & 0 \\
 1 & 0 & 0 & 0 \\
 0 & 0 & 0 & 1 \\
 \end{pmatrix},\ 
  {\mathrm{\bf M}_{24}} = 
\begin{pmatrix} 
 1 & 0 & 0 & 0 \\
 0 & 0 & 0 & 1 \\
 0 & 0 & 1 & 0 \\
 0 & 1 & 0 & 0 \\
 \end{pmatrix},
\end{eqnarray}
where $\mathrm{\bf M}_{ij}$ represents the mutual exchange of states $i$ and $j$ while the other two states are fixed. These holonomy matrices, describing the stroboscopic encircling of EP2, are 2-cycle permutations called transpositions. In cyclic notation, they can be written in simpler symbolic forms: $(12)$, $(23)$, $(34)$, $(14)$, $(13)$, and $(24)$, respectively. We emphasize that the stroboscopic encircling of {\it any multiple EP2s} is equivalent to the product of the holonomy matrices. Also, the stroboscopic encircling of {\it any higher-order EPs} can be described as a product of the holonomy matrices because every permutation can be written as a product of transpositions. Note that the sequences of the holonomy matrix products can be defined conveniently as the branch cut crossing sequences along the encircling trajectories~\cite{Ryu12, Zho18}.

{\it Encircling multiple EPs with a single loop -} Here, we classify the stroboscopic encircling of multiple EP2s by analyzing the eigenvalue structures of the products of the holonomy matrices. Let us begin with considering a single EP2 case. A single EP2 corresponds to one of six holonomy matrices, $M_{12}$, $M_{23}$, $M_{34}$, $M_{14}$, $M_{13}$, or $M_{24}$. All of the holonomy matrices have the same eigenvalues $(-1,1,1,1)$, which are the solutions of the secular equation $(\lambda^2 - 1)(\lambda - 1)^2 = 0$. The eigenvalues or secular equation implies that two states exchange their eigenvalues after encircling EP2 once [$(\lambda^2-1)$-factor in the secular equation], while the other two states remain as they are [$(\lambda-1)$-factor in the secular equation for each state]. In the view of the permutation group, all six holonomy matrices are in the same conjugacy class. The conjugacy class of $S_N$ is denoted by a cyclic structure with the notation $(1^{\alpha_1} 2^{\alpha_2} \cdots N^{\alpha_N})$, where $\sum_{n=1}^N n\alpha_n=N$, $\alpha_n$ are non-negative integers~\cite{Ham62}. So the conjugacy class that $M_{ij}$s reside in is $(1^2 2^1)$, where $\alpha_i=0$ terms are neglected for simplicity.

\begin{table*}
\caption{Number of 2nd order EPs (NEP), classes, corresponding secular equations, eigenvalues of the product of holonomy matrices, number of possible permutations (NOP), and the product of holonomy matrices and outcome permutations of the examples in Fig.~\ref{meps} for the four-state case. Asterisks mark the multiple EPs that do not correspond to the 4th order EP.}
\begin{center}
\renewcommand{\arraystretch}{1.1}
\scalebox{1.0}{
\begin{tabular}{|c|l|l|c|r|l|} 

\toprule
 \thead{NEP}    & \thead{Class} & \thead{Secular Equation} & \thead{Eigenvalues} & \thead{NOP} & \thead{Examples of Fig.~\ref{meps}}\\

\toprule
~1* & $(1^2 2^1)$  & $(\lambda^2 -1)(\lambda -1)^2 = 0$ & $(-1,1,1,1)$ & $6$ & $\mathrm{\bf M}_{12}$, (12) \\
\hline
~2* & $(1^1 3^1)$  & $(\lambda^3 - 1)(\lambda - 1) = 0$ & $((-1)^{2/3},-(-1)^{1/3},1,1)$ & $24$ & $\mathrm{\bf M}_{13}\mathrm{\bf M}_{12}$, (123) \\
\hline
2& $(2^2)$ & $(\lambda^2 - 1)^2 = 0$ & $(-1,-1,1,1)$ & $6$ & $\mathrm{\bf M}_{34}\mathrm{\bf M}_{12}$, (12)(34)\\
\hline
3 & $(4^1)$   & $(\lambda^4 - 1) = 0$ & $(-1,i,-i,1)$ & $96$ & $\mathrm{\bf M}_{34}\mathrm{\bf M}_{23}\mathrm{\bf M}_{12}$, (1432)\\
\hline
~3*& $(1^2 2^1)$   & $(\lambda^2 -1)(\lambda -1)^2 = 0$ & $(-1,1,1,1)$ & $24$ & $\mathrm{\bf M}_{13}\mathrm{\bf M}_{23}\mathrm{\bf M}_{12}$, (23) \\
\hline
4& $(1^1 3^1)$     & $(\lambda^3 - 1)(\lambda - 1) = 0$ & $((-1)^{2/3},-(-1)^{1/3},1,1)$ & $216$ & $\mathrm{\bf M}_{14}\mathrm{\bf M}_{34}\mathrm{\bf M}_{23}\mathrm{\bf M}_{12}$, (243) \\
\hline
4& $(2^2)$  & $(\lambda^2 - 1)^2 = 0$ & $(-1,-1,1,1)$ & $144$ & $\mathrm{\bf M}_{13}\mathrm{\bf M}_{34}\mathrm{\bf M}_{23}\mathrm{\bf M}_{12}$, (14)(23) \\
\hline
5& $(4^1)$     & $(\lambda^4 - 1) = 0$ & $(-1,i,-i,1)$ & $360$ &$\mathrm{\bf M}_{13}\mathrm{\bf M}_{14}\mathrm{\bf M}_{34}\mathrm{\bf M}_{23}\mathrm{\bf M}_{12}$, (1324) \\
\hline
5& $(1^2 2^1)$    & $(\lambda^2 -1)(\lambda -1)^2 = 0$ & $(-1,1,1,1)$ & $360$ & $\mathrm{\bf M}_{24}\mathrm{\bf M}_{14}\mathrm{\bf M}_{34}\mathrm{\bf M}_{23}\mathrm{\bf M}_{12}$, (34) \\
\hline
6& $(1^1 3^1)$     & $(\lambda^3 - 1)(\lambda - 1) = 0$ & $((-1)^{2/3},-(-1)^{1/3},1,1)$ & $432$ & $\mathrm{\bf M}_{24}\mathrm{\bf M}_{13}\mathrm{\bf M}_{14}\mathrm{\bf M}_{34}\mathrm{\bf M}_{23}\mathrm{\bf M}_{12}$, (134)\\
\hline
6& $(2^2)$  & $(\lambda^2 - 1)^2 = 0$ & $(-1,-1,1,1)$ & $192$ & $\mathrm{\bf M}_{24}\mathrm{\bf M}_{13}\mathrm{\bf M}_{34}\mathrm{\bf M}_{14}\mathrm{\bf M}_{23}\mathrm{\bf M}_{12}$, (12)(34) \\
\hline
6& $(1^4)$     & $(\lambda - 1)^4 = 0$ & $(1,1,1,1)$ & $96$ & $\mathrm{\bf M}_{34}\mathrm{\bf M}_{24}\mathrm{\bf M}_{13}\mathrm{\bf M}_{14}\mathrm{\bf M}_{23}\mathrm{\bf M}_{12}$, (1)\\
\hline
\end{tabular}}
\end{center}
\label{table_4}
\end{table*}

Now, when one more EP2 is added, two different kinds of eigenvalue structures emerge. First, two EP2s can share a single state, e.g., $M_{12}$ and $M_{13}$ [see the first diagram of 2-EP in Fig.~\ref{meps}]. The secular equation for this case is $(\lambda^3 - 1)(\lambda - 1)=0$, which implies that three states undergo the cyclic change [($\lambda^3 - 1$)-factor in the equation] while the remaining state is unchanged [($\lambda - 1$) factor in the equation]. There are $24$ kinds of holonomy matrices in this case, and the conjugacy class is $(1^1 3^1)$. The other case is two EP2s changing disjoint states, e.g., $M_{12}$ and $M_{34}$ [see the second diagram of 2-EP in Fig.~\ref{meps}]. The secular equation is $(\lambda^2 - 1)^2 = 0$, which implies that two states are exchanged while two remain. Here, there are $6$ kinds of holonomy matrices, and the conjugacy class is $(2^2)$.

If we consider three EP2s, we have two different conjugacy classes: first, all four states link via the three EP2s, and second, only three states are interconnected through the three EP2s while the one remaining state is isolated. For the first case, Fig.~\ref{meps} examines the permutation of $\mathrm{M}_{34}\mathrm{M}_{23}\mathrm{M}_{12}$ as an example, where the product undergoes a cyclic exchange of all four states. There are $96$ holonomy matrices of this kind, and the conjugacy class is $(4^1)$. The 4-cycle structure corresponds to the roots of the multi-valued function of $z^{1/4}$ when sufficiently far from the EP2s. The second case is equivalent to the case of three EP2s in a closed three-state system accompanied by a single independent state, e.g., the permutation $(23)$ of $\mathrm{M}_{13}\mathrm{M}_{23}\mathrm{M}_{12}$. We note here that only multiple EP2s stemming from a single EP4 are regarded in this work, though other associated classes also follow our classification. That is, we are not interested in 1-EP, the first diagram of 2-EP, or the second diagram of 3-EP in Fig.~\ref{meps} and Table~\ref{table_4} since they are not equivalent to the case of multiple EP2s split from a EP4.

When more than four EPs are involved, every state links, as we can see in Fig.~\ref{meps}. Concerning the constituted number of EP2s, the class, secular equation, eigenvalue, and number of permutations are summarized in Table~\ref{table_4}. Surprisingly, a total of 1956 possible permutations of the transposition products converge into only five classes, as below:
\begin{align}
\label{sols}
\begin{array}{lll}
(4^1) &: (\lambda^4 - 1) = 0, & (-1,i, -i, 1), \\
(1^1 3^1) &: (\lambda^3 - 1)(\lambda - 1) = 0, & ((-1)^{\frac{2}{3}}, -(-1)^{\frac{1}{3}},1,1), \\
(2^2) &: (\lambda^2 - 1)^2 = 0, &(-1,-1,1,1), \\
(1^2 2^1) &: (\lambda^2 - 1)(\lambda-1)^2 = 0, &(-1,1,1,1), \\
(1^4) &: (\lambda - 1)^4 = 0, &(1,1,1,1).
\end{array}
\end{align}
Here, we emphasize that our classification of the stroboscopic encircling into conjugacy classes of a permutation group is not restricted to the four-state case but is consistently valid for an arbitrary number of states (see Supplementary Information for details).

In this work, we used six holonomy matrices only once during the encircling of multiple EP2s to reduce the possible number of permutations. Our classification scheme is still valid for the repeated use of holonomy matrices. Also, note that while the equivalence between the loops and more complicated permutations can be well understood in terms of homotopy~\cite{Zho18}, it seems rather detailed. The conjugacy class, on the other hand, is more reliable. For a given encircling curve and branch cuts of EPs, the stroboscopic encircling may depend on the starting state, while the conjugacy class does not. Also, the conjugacy class is independent of the branch cuts when the EPs outside the encircling curve are irrelevant. Here, an irrelevant EP means the case when the branch cut of an EP can be pulled out to the encircling curve without crossing the other EPs (see Supplementary Information for details). Consequently, we can safely use arbitrarily defined branch cuts in our classification procedures, assuming that the encircling curve is large enough to make EPs outside of the curve irrelevant.

\begin{figure*}
\begin{center}
\includegraphics[width=0.8\textwidth]{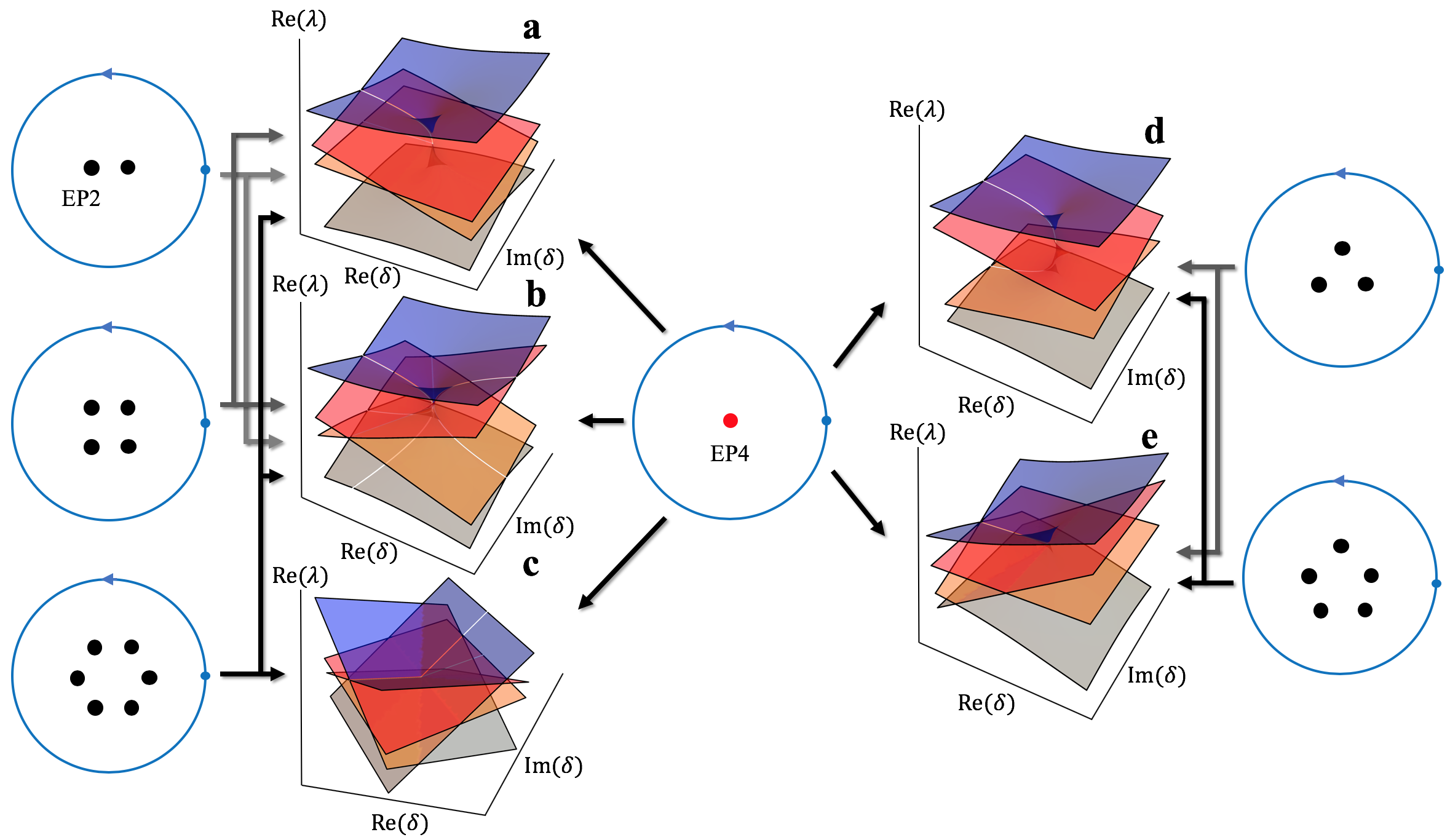}
\caption{Classifications of multiple 2nd order EPs associated with a 4th order EP. The stroboscopic encircling of the 4th order EP (EP4, red dot) is based on a starting point (blue dot) in counterclockwise direction (blue arrows). {\bf a}--{\bf e}. Riemann surfaces of the real parts of the eigenvalues around EP4 with {\bf a}: cubic-root and linear branches with permutation $(143)$ in conjugacy class $(1^1 3^1)$, {\bf b}: a pair of square-root branches, $(14)(23)$ in class $(2^2)$, {\bf c}: linear branches, the identity $(1)$ in class $(1^4)$ ({\bf c}), {\bf d}: a quartic-root branch, $(1342)$ in class $(4^1)$, and {\bf e}: square-root and a pair of linear branches, $(14)$ in class $(1^2 2^1)$. The four branches 1, 2, 3, and 4 (blue, red, orange, and gray surfaces) are defined by the order of the real parts of the eigenvalues. (left) Two, four, and six 2nd order EPs (black dots) and stroboscopic encircling based on a fixed starting point (blue dots) in counterclockwise direction (blue arrows). Two, four, and six second-order EPs can generate Riemann surfaces with cubic-root and linear branches, a pair of square-root branches, and four linear branches. (right) Three and five second-order EPs, which can generate Riemann surfaces with the same topological structures. Three and five 2nd order EPs can generate Riemann surfaces with both quartic-root and square-root branches and a pair of linear branches.
}
\label{intro}
\end{center}
\end{figure*}

{\it Encircling a 4th order EP -} Lastly, we consider a $4\times 4$ Hamiltonian $H$,
\begin{eqnarray}
H &=& H_0 + \delta H_p\ ,
\label{pertH}
\end{eqnarray}
where $H_0$ is the $4\times4$ Jordan normal form, $H_p$ is a perturbation, and $\delta$ is a complex number corresponding to a two-dimensional parameter space for Riemann surfaces. The EP4 is located at $\delta = 0$. Depending on the form of $H_p$, the Riemann surfaces are classified into the five classes in Eq. (\ref{sols}). We examine the perturbations that generate Riemann surfaces of each class as follows:
\begin{align}
\label{Jordan}
{H_{p}^{(4^1)}} &=
\begin{pmatrix} 
 0 & 0 & 0 & 0 \\
 0 & 0 & 0 & 0 \\
 0 & 0 & 0 & 0 \\
 1 & 0 & 0 & 0 \\
 \end{pmatrix}, \
{H_{p}^{(1^1 3^1)}} =
\begin{pmatrix} 
 0 & 0 & 0 & 0 \\
 0 & 0 & 0 & 0 \\
 1 & 0 & 0 & 0 \\
 0 & 1 & 0 & 0 \\
 \end{pmatrix},\
 {H_{p}^{(2^2)}} =
\begin{pmatrix} 
 0 & 0 & 0 & 0 \\
 1 & 0 & 0 & 0 \\
 0 & -1 & 0 & 0 \\
 0 & 0 & 1 & 0 \\
 \end{pmatrix}, \nonumber\\
 {H_{p}^{(1^22^1)}} &= 
\begin{pmatrix} 
 1 & 0 & 0 & 0 \\
 0 & 0 & 0 & 0 \\
 0 & 1 & 0 & 0 \\
 0 & 0 & 1 & 0 \\
 \end{pmatrix},\
  {H_{p}^{(1^4)}} =
\begin{pmatrix} 
 0 & 0 & 0 & 0 \\
 0 & 0 & 0 & 1 \\
 0 & -1 & 0 & 0 \\
 0 & 0 & 1 & 0 \\
 \end{pmatrix}.
\end{align}
Figure \ref{intro}{\bf a} shows the Riemann surface of the real part of the eigenvalues with $H_p = H_p^{(1^13^1)}$ in parameter space $[\mathrm{Re}(\delta),\ \mathrm{Im}(\delta)]$. As the notation already indicates, the surface corresponds to the conjugacy class $(1^13^1)$. To illustrate this, let us follow the branches of the eigenvalues while encircling EP4. Starting from the rightmost corner in the uppermost sheet, state 1 (blue surface), the branches after the encircling of EP4 are
\begin{align}
\text{state 1 (blue surface)} \mapsto \text{state 4 (gray surface)} \nonumber \\
\mapsto \text{state 3 (orange surface)} \mapsto \text{state 1 (blue surface)}. \nonumber
\end{align}
Here, ``$\mapsto$'' stands for the one-time counterclockwise encircling of EP4. When starting from the remaining state 2 (red surface), one can easily see that the state does not change:
\begin{align}
\text{state 2 (red surface)} \mapsto \text{state 2 (red surface)}. \nonumber
\end{align}
So the stroboscopic encircling of the Riemann surface is the permutation $(143)$, which is in the conjugacy class $(1^1 3^1)$. The other surfaces in Fig. \ref{intro} can be analyzed by the same procedure. The Riemann surfaces in Fig.~\ref{intro}{\bf b--e} are the real parts of the eigenvalues with $H_p$ $=$ $H_p^{(2^2)}$, $H_p^{(1^4)}$, $H_p^{(4^1)}$, $H_p^{(1^2 2^1)}$, respectively. The corresponding stroboscopic encirclings are $(14)(23)$, $(1)$, $(1342)$, and $(14)$, such that they are in the conjugacy classes indicated in the perturbation.

When small symmetry-breaking disorders are introduced in $H_0$, EP4 is split into multiple EP2s around the original position of EP4. Although the structure of the Riemann surface near the multiple EP2s might be changed significantly, the structure at a far distance is still preserved and the associated conjugacy class does not change. Note that we separate disorders and perturbations for the simplicity of discussion, though they are not independent in general. For example, we can set the former as a boundary shape distortion in deformed optical microcavities, while the latter can be any two system parameters, including the boundary shape.

{\it Parity of permutation -} We remark that the correspondence between the EP2s and transpositions shows that the conjugacy classes in Eq.~(\ref{sols}) can be classified by the parity of the number of included EP2s. The first [$(4^1)$] and the fourth [$(1^2 2^1)$] have odd parity as three and five EP2s embody their permutations, respectively. On the other hand, the second $(1^1 3^1)$], the third [$(2^2)$], and the fifth [$(1^4)$] have even parity since two, four, and six EP2s are involved, respectively (see Table~\ref{table_4}). Figure~\ref{intro} explicitly demonstrates this EP2-dependent classification of permutations that are associated with the diverse topological structures of the Riemann surfaces. Note that the depicted Riemann surfaces are the consequences of the split odd or even number of EP2s from EP4 by the continuous increase of the disorder.

\subsection{Permutation as the product of disjoint cycles}

\begin{figure*}
\begin{center}
\includegraphics[width=.7\textwidth]{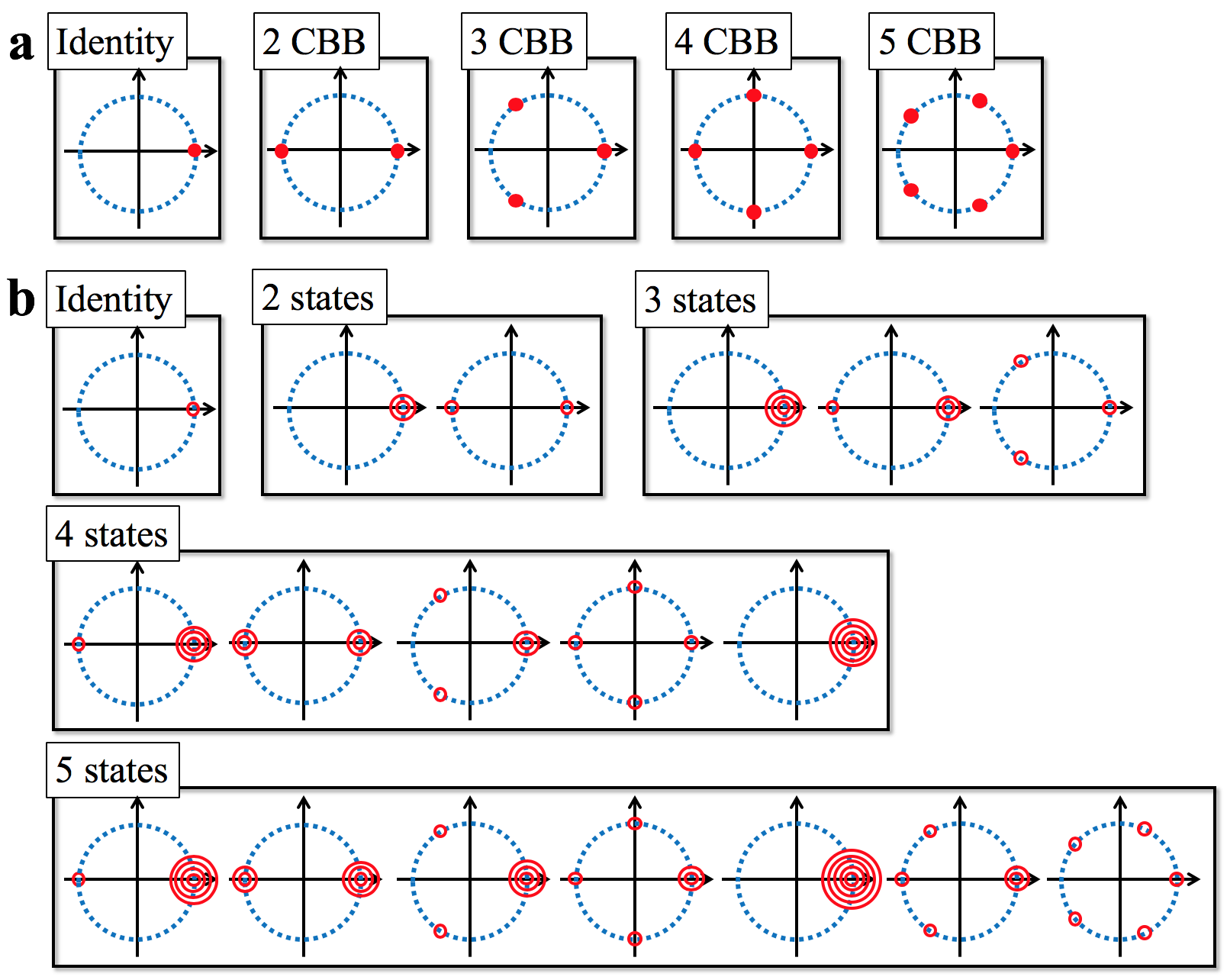}
\caption{{\bf a.} Cyclic building blocks. Complex eigenvalues of N-cycle building blocks for N states in complex planes [$\mathrm{Re}(\lambda)$, $\mathrm{Im}(\lambda)$]. Identity: $(1)$ of the 1-cycle building block for one state, 2 CBB: $(-1,1)$ of the 2-cycle building block for two states, 3 CBB: $((-1)^{2/3},-(-1)^{1/3},1)$ of the 3-cycle building block for three states, 4 CBB: $(-1,i,-i,1)$ of the 4-cycle building block for four states, and 5 CBB: $((-1)^{4/5},(-1)^{2/5},-(-1)^{1/5},-1(-1)^{3/5})$ of the 5-cycle building block for five states. We use the following symbols: Identity \textcircled{1}, 2 CBB \textcircled{2}, 3 CBB \textcircled{3}, 4 CBB \textcircled{4}, and 5 CBB \textcircled{5}.
{\bf b.} Reassembled complex eigenvalues of the products of holonomy matrices for one state (identity), two states, three states, four states, and five states, respectively. The number of circles in a concentric circle represents the number of the same eigenvalues. The symbolic expressions are as follows. Identity: \textcircled{1}; 2 states: \textcircled{1} + \textcircled{1}, \textcircled{2}; 3 states: \textcircled{1} + \textcircled{1} + \textcircled{1}, \textcircled{1} + \textcircled{2}, \textcircled{3}; 4 states: \textcircled{1} + \textcircled{1} + \textcircled{2}, \textcircled{2} + \textcircled{2}, \textcircled{1} + \textcircled{3}, \textcircled{4}, \textcircled{1} + \textcircled{1} + \textcircled{1} + \textcircled{1}; 5 states: \textcircled{1} + \textcircled{1} + \textcircled{1} + \textcircled{2}, \textcircled{1} + \textcircled{2} + \textcircled{2}, \textcircled{1} + \textcircled{1} + \textcircled{3}, \textcircled{1} + \textcircled{4}, \textcircled{1} + \textcircled{1} + \textcircled{1} + \textcircled{1} + \textcircled{1}, \textcircled{2} + \textcircled{3}, \textcircled{5}.}
\label{bb}
\end{center}
\end{figure*}

{\it Any permutation can be expressed as the product of disjoint cycles -} It is instructive to see that any kind of permutation can be constructed by repetitions of simple $n$-cyclic building blocks ($n$ CBB), as shown in Fig.~\ref{bb}{\bf a}. Each of these building blocks corresponds to the longest-cyclic holonomy matrix taking $n$ states into account.

We define the symbol \textcircled{n} to represent the set of $n$ states that pass through the encirclings of EPs before returning to the starting state. Consider an $N$ state system. The $N$ states can be split into subsets where the states within a subset can be reached by multiple stroboscopic encirclings while other subsets cannot. This subset is an $n$-cycle, where $n$ is the number of states in the subset. The number of possibilities of splitting $N$ states is the number of conjugacy classes. So the conjugacy class can be indexed by the number of cycles, denoted as the sum of \textcircled{n}s. For example, in a four-state system, there can be one $2$-cycle and two $1$-cycle, which is denoted by \textcircled{1}+\textcircled{1}+\textcircled{2}. This is the class $(1^2 2^1)$ in the notation of the previous section.

Although all the stroboscopic encirclings are classified by the conjugacy class, there are two cases in which the conjugacy classes are not realized in the stroboscopic encirclings: the identity class in $N=2$ and $N=3$ states. The reason why there is no identity permutation in the two- and three-state cases is as follows. We consider the permutations of set $\{1, 2, 3,\cdots, N\}$. When $N > m$, we cannot make an identity permutation using the products of $m$ different transpositions, i.e., $m$ different EP2s, if we assume that all elements of the set have to be changed by permutations at least one time. There is no identity since $N=2$ and $m=1$ for the two-state case, and $N=3$ and $m=3$ for the three-state case, with the identity an even permutation. If we use the same holonomy matrices multiple times during encircling, the identities can exist even in the cases of two and three states.

This construction of classes can be done by using the holonomy matrices. In Fig. \ref{bb}{\bf a}, we plot solutions of the equation $\lambda^n-1=0$ where $n=1, 2, 3, 4, 5$, which are the factors in the secular equations of the holonomy matrices shown in Table \ref{table_4}. These correspond to the $n$ CBB. The secular equations are obtained by multiplying $n$ CBBs to make an $N$th order equation in $\lambda$. For example, the third diagram in the four-state case in Fig. \ref{bb}{\bf b} represents the secular equation $(\lambda^3-1)(\lambda-1)=0$.

Utilizing the cyclic building blocks, all possible classes of complicated topological structures associated with single or multiple arbitrary-order EPs in cases with an arbitrary number of states can be revealed. For example, 5th, 6th, and 7th order EPs exhibit 7, 11, and 15 classes, respectively (see Fig.~\ref{bb} and Supplementary Information). The number of possible topologically distinct classes for $N$ states, which is the same as the possible number of distinct Riemann surfaces with $N$ surfaces around EPNs or multiple EPMs ($M < N$), is $2, 3, 5, 7, 11, \cdots$, for $N = 2, 3, 4, 5, 6, \cdots$, respectively \cite{oeis}. 

This classification can also help search for higher-order EPs. For example, if we know that the permutation class is $(1^1 3^1)$ for a four-state system, it is guaranteed to find a $3$rd order EP or numerous EPs behaving as $3$rd order EPs far from the EPs. However, when the permutation class is $(2^2)$, it is not guaranteed to find a $3$rd order EP, although the possibility is not excluded in principle.

\subsection{Realization of multiple EPs - Microcavities}

\begin{figure*}[htp]
\begin{center}
\includegraphics[width=\textwidth]{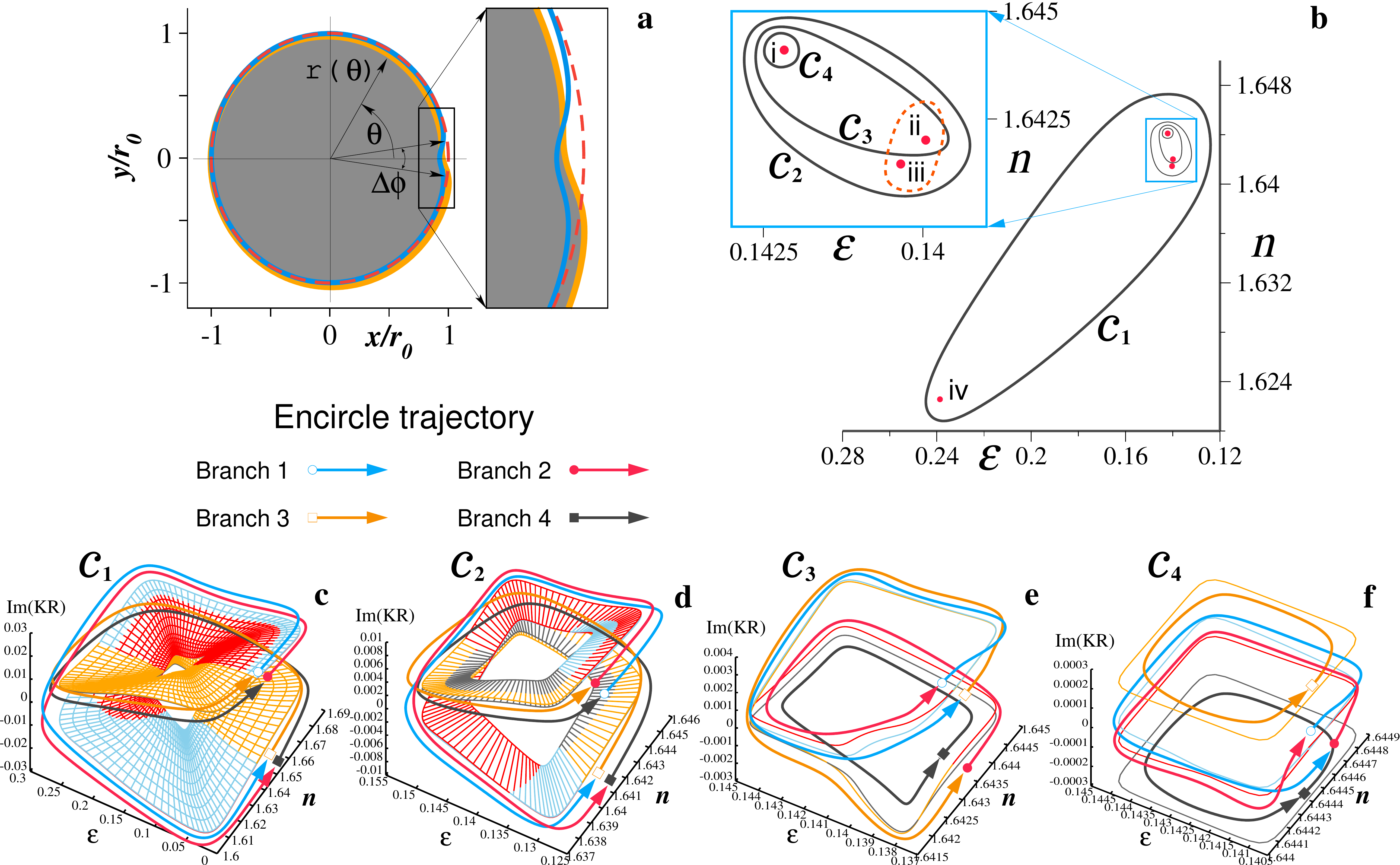}
\caption{{\bf a.} Deformed microcavity exemplifying three different symmetric boundary shapes: rotational-symmetric circular (red-dashed), mirror-symmetric deformed (blue-solid), and asymmetric deformed (orange-solid). The shaded (white) inside (outside) region indicates the cavity domain (vacuum) having a refractive index of $n>1$ $(n=1)$. {\bf b.} Four EP2s (red dots) and contour loops in $(n,
\varepsilon)$ parameter space enclosing four ($C_1$), three ($C_2$), two ($C_3$), and one ($C_4$) EP2s. {\bf c}--{\bf f}. Riemann surfaces of \imag{KR} whose outer edges correspond to the four contour loops $C_j$ in \tbf{b}. Here, we show the relative wavenumbers with respect to the average of the four wavenumbers as $KR=kR-1/4\sum_{j=1}^4k_jR$. The colored arrows stand for the four branches along the stroboscopic encircling around EPs. The stroboscopic encirclings represent \tbf{c} (13)(24) for $C_1$, \tbf{d} (1324) for $C_2$, \tbf{e} (132) for $C_3$, and \tbf{f} (12) for $C_4$.}
\label{fig_mc}
\end{center}
\end{figure*}

To benchmark the validity of our argument in terms of realistic physical systems, we demonstrate a numerical experiment considering a two-dimensional asymmetric deformed optical microcavity. The microcavity consists of extinction-free dielectric material defined inside a non-circular boundary shape with no rotational or mirror-reflection symmetry; namely, an averaged-two-notch spiral in which the two notches face one another in the opposite direction, creating a single concave part between them as shown in Fig.~\ref{fig_mc}{\bf a} (see the Method section). To generate as many EP2s as possible for a given finite number of modes, the boundary shape is multi-valued-parameterized. By fine-tuning the parameters, we achieve multiple EP2s in the small range of parameter space spanned by shape parameter $\varepsilon$ and refractive index $n$, while other parameters are fixed.

\begin{figure*}[t]
\begin{center}
\includegraphics[width=\textwidth]{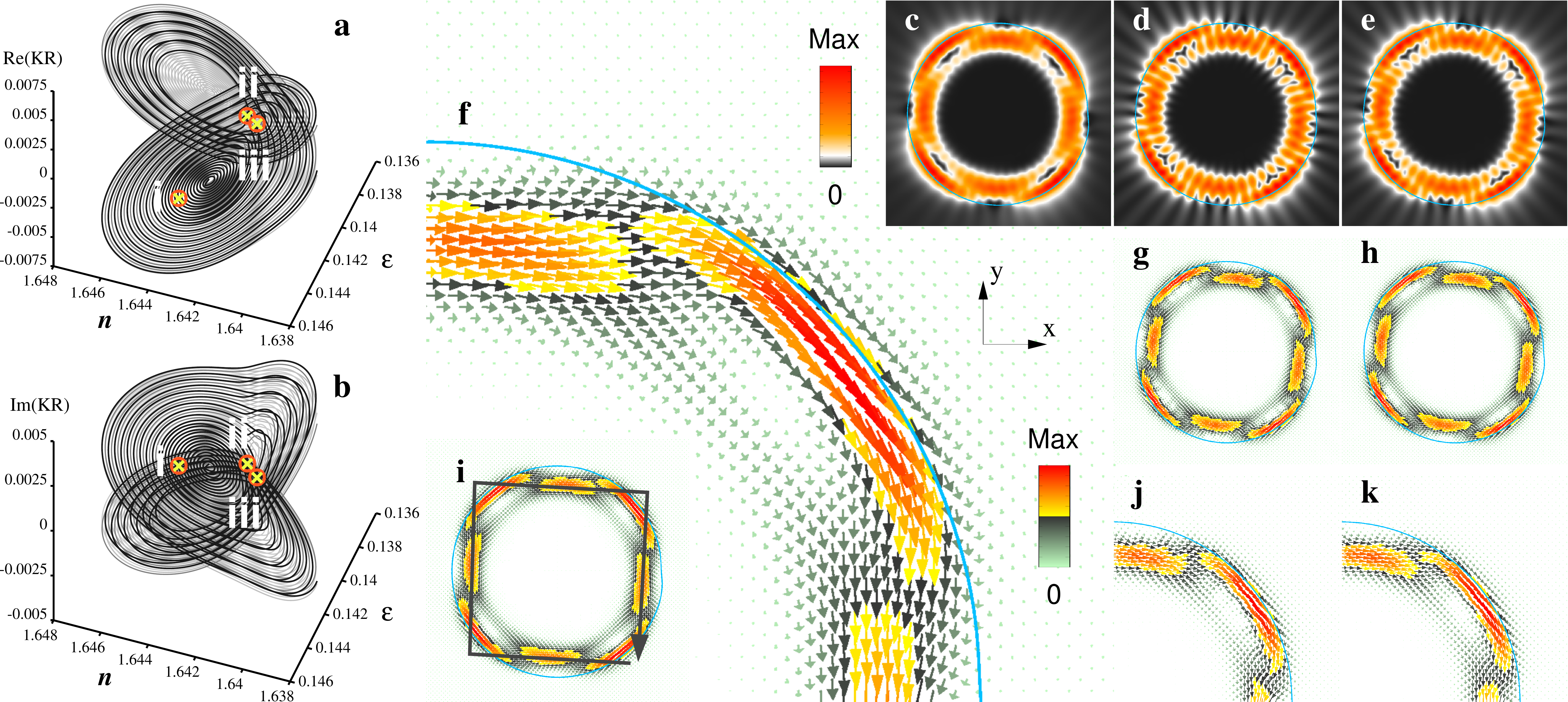}
\caption{{\bf a}. Real and \tbf{b}. imaginary  parts of $KR$s for three EPs \{\texthelv{i, ii, iii}\} enclosed by $C_2$ shown in Fig.~\ref{fig_mc}. The intensity ($|\psi|^2$) and Poynting vector [$\vec{S}\equiv\alpha \text{Im}(\psi^*\nabla \psi)$] of the mode at EP-\texthelv{i} are shown in \tbf{c} and \tbf{i}, respectively, and magnified in \tbf{f}. In the same order, those of EP-\texthelv{ii} and EP-\texthelv{iii} are in (\tbf{d,g,j}) and in (\tbf{e,h,k}), respectively. Here, the normalization constant $\alpha$ is introduced to scale the largest $||\vec{S}||$ to unity. All the examined modes at the EPs in our asymmetric deformed optical microcavity are localized on the classical periodic orbit of a square geometry. At the same time, they exhibit a preferred propagation direction toward the clockwise direction, as we can see by following the exemplified square-shaped polygonal arrow in \tbf{i}.}
\label{fig_mc2}
\end{center}
\end{figure*}

\begin{table*}[t]
\centering
\caption{Numerically obtained wavenumbers of the EPs shown in Fig.~\ref{fig_mc}. The pairwise values in the shaded cells are of EPs obtained with a numerical precision of $\sim 10^{-5}$. The index $j$ is defined in the Riemann surfaces of Fig.~\ref{fig_mc}.}
\label{tab:myt}
\renewcommand{\arraystretch}{1.5}
\setlength{\tabcolsep}{5pt}
{
\begin{tabular}{ccccccccc}

\toprule
  &  \multicolumn{2}{c}{EP-\texthelv{i}} &  \multicolumn{2}{c}{EP-\texthelv{ii}} &  \multicolumn{2}{c}{EP-\texthelv{iii}} &  \multicolumn{2}{c}{EP-\texthelv{iv}}\\ 
  
 \toprule
 \multicolumn{1}{|c"}{$(n,\ \varepsilon)$} &  \multicolumn{2}{c"}{(1.6441,\ 0.1423)} &  \multicolumn{2}{c"}{(1.6420,\ 0.1400)} &  \multicolumn{2}{c"}{(1.6415,\ 0.1403)} &  \multicolumn{2}{c|}{(1.6226,\ 0.2387)}\\
  \hline

\multicolumn{1}{|c"}{index $j$} &\multicolumn{1}{c|}{Re$(kR)$} &  \multicolumn{1}{c"}{Im$(kR)$} &  \multicolumn{1}{c|}{Re$(kR)$} &  \multicolumn{1}{c"}{Im$(kR)$} &  \multicolumn{1}{c|}{Re$(kR)$} &  \multicolumn{1}{c"}{Im$(kR)$} &  \multicolumn{1}{c|}{Re$(kR)$} &  \multicolumn{1}{c|}{Im$(kR)$} \\

\toprule

\multicolumn{1}{|c"}{1} &\multicolumn{1}{c|}{\cellcolor{lightgray!50}17.1979} &  \multicolumn{1}{c"}{\cellcolor{lightgray!50}$-$0.01947} &  \multicolumn{1}{c|}{17.2251} &  \multicolumn{1}{c"}{$-$0.02227} & \multicolumn{1}{c|}{\cellcolor{lightgray!50}17.2289} & \multicolumn{1}{c"}{\cellcolor{lightgray!50}$-$0.02136} &  \multicolumn{1}{c|}{17.4355} &  
\multicolumn{1}{c|}{$-$0.02364}\\ \hline

\multicolumn{1}{|c"}{2} &\multicolumn{1}{c|}{\cellcolor{lightgray!50}17.1979} &  \multicolumn{1}{c"}{\cellcolor{lightgray!50}$-$0.01946} &  \multicolumn{1}{c|}{\cellcolor{lightgray!50}17.2232} &  \multicolumn{1}{c"}{\cellcolor{lightgray!50}$-$0.02104} & \multicolumn{1}{c|}{17.2297} & \multicolumn{1}{c"}{$-$0.01944} &  \multicolumn{1}{c|}{17.4353} &  \multicolumn{1}{c|}{$-$0.02312}\\
\hline

\multicolumn{1}{|c"}{3} &\multicolumn{1}{c|}{17.2079} &  \multicolumn{1}{c"}{$-$0.02187} &  \multicolumn{1}{c|}{\cellcolor{lightgray!50}17.2232} &  \multicolumn{1}{c"}{\cellcolor{lightgray!50}$-$0.02106} & \multicolumn{1}{c|}{17.2277} & \multicolumn{1}{c"}{$-$0.02288} &  \multicolumn{1}{c|}{\cellcolor{lightgray!50}17.3942} &  \multicolumn{1}{c|}{\cellcolor{lightgray!50}$-$0.03111}\\
\hline

\multicolumn{1}{|c"}{4} &\multicolumn{1}{c|}{17.2083} &  \multicolumn{1}{c"}{$-$0.02240} &  \multicolumn{1}{c|}{17.2217} &  \multicolumn{1}{c"}{$-$0.02026} & \multicolumn{1}{c|}{\cellcolor{lightgray!50}17.2289} & \multicolumn{1}{c"}{\cellcolor{lightgray!50}$-$0.02138} &  \multicolumn{1}{c|}{\cellcolor{lightgray!50}17.3942} &  \multicolumn{1}{c|}{\cellcolor{lightgray!50}$-$0.03111}\\
\hline
\end{tabular}
}
\end{table*}

According to the selection rule for coupling~\cite{ep-prl,sgol,Kul18,Kul19}, we can induce even- and odd-symmetric paired modes (i.e., four modes are involved) at the EP2s in a mirror-reflection symmetric system. We begin with a quasi-degenerate pair of modes, $k_{l,m}R=k_{1,24}R=17.256-i2.242\times 10^{-4}$ and $k_{2,20}R=17.256-i3.743\times 10^{-2}$, in a circular cavity, of which $\text{Re}(k_{1,24}R)=\text{Re}(k_{2,20}R)$ when refractive index $n=1.639$. As these modes are associated with the Fermi resonance relation $(\delta l,\delta m)=(|1-2|,|24-20|)=(1,4)$, we can infer that the EPs formed by these pairs would have their classical counterpart periodic orbits~\cite{fermi,ep-prl} characterized by a winding number $(1,4)$, i.e., a square orbit that bounces four times for one rotation.

Along with those of the selection rule-based EP formations, there is another type of EP formation mechanism that functions by the asymmetric backscattering process. The essential difference between the two types is that while the former case coalesces the different $m$ modes to form EPs under the same mirror-reflection parity (i.e., even or odd), in the latter case, the same $m$ modes having different mirror parities merge to form EPs. Typically, EPs from the first case manifest a local vortex mode structure localized on the classical periodic orbits, whereas the second one exhibits a global chirality of the mode propagation direction (e.g.,~\cite{sgol,ep-prl}).

We can expect that if the two mechanisms operate simultaneously, multiple EP2s associated with only four targeted modes can be observable at once in a sufficiently small parameter range. Moreover, if this is the case, it can be anticipated that these EPs will be such a mode that is localized on the periodic orbits exhibiting a single directional propagation. Figure~\ref{fig_mc}\tbf{b} shows a group of four EPs associated with the modes $k_{1,24}R$ and $k_{2,20}R$ in the circular cavity. The inter-connectivity of the Riemann surfaces, which verifies that these EPs indeed are of the modes $k_{1,24}R$ and $k_{2,20}R$, is clearly demonstrated in Fig.~\ref{fig_mc}\tbf{c}--\tbf{f}. As expected, all the considered modes here reveal a clear localization on the 4-period classical orbit and a unidirectional propagation direction. Note that the over-penetrated corner positions of the square-shape in Fig.~\ref{fig_mc2}\tbf{i} are related to the Goos--H\"{a}nchen shift effect~\cite{goos}. In the figure, scaled relative wavenumbers,
\begin{align}
    KR=kR-\frac{1}{4}\sum_{j=1}^4k_jR\ ,
\end{align}
are used for better visibility. Since the two-dimensional Helmholtz wave equation cannot be solved analytically for the non-separable boundary shape of microcavities, we employ the boundary element method (BEM)~\cite{bem} to numerically solve the problem. The original values of $k_jR$ at the EPs obtained via BEM are summarized in Table~\ref{tab:myt}.

We confirm that our argument of the classification and the permutation over the branch switching scenario for encircling multiple EPs is consistently valid in the actual physical system. The Riemann surfaces in Fig.~\ref{fig_mc}\tbf{c}--\tbf{f} account for the permutation sequences of the eigenstate interchanges brought by the four encircled EPs in our microcavities. Details of the precise four-mode-tracing strategy for numerically encircling multiple EPs are given in the Method section. As the branch classification does not depend on the choice of branch cuts, an arbitrary refractive index value of $n\sim 1.644$ is chosen to assign the identification numbers of the permutation elements (i.e., distinguishing marker numbers of the branches) to the eigenstates. Note that this arbitrary choice of the branch cut has no spoiling effect at all on our encircling permutation, and the conventional branch cuts, e.g., the same real or imaginary parts of the eigenvalues, are unnecessary. By tracing out the arrowed encircling contours $C_j$ in Fig.~\ref{fig_mc}\tbf{c}--\tbf{f}, we can identify the branch switchings:
\begin{align}
\begin{matrix}
 C_4 &\Longrightarrow& (12),\quad   &C_3 &\Longrightarrow& (132)\\
 C_2 &\Longrightarrow& (1324),\quad &C_1 &\Longrightarrow& (13)(24)
 \end{matrix}    
 \ .
\end{align}
Straightforwardly, we can infer that these combined permutations are equivalent to the cases when the encircling of each EP2 corresponds to the holonomy matrix operation:
\begin{align}
\begin{matrix}
\begin{tikzcd}\scalebox{1.5}{$\circlearrowleft$}\end{tikzcd}\text{EP-}\texthelv{\ i\ }&\iff& \mathrm{\bf M}_{12} \\
\begin{tikzcd}\scalebox{1.5}{$\circlearrowleft$}\end{tikzcd}\text{EP-}\texthelv{ii\ }&\iff& \mathrm{\bf M}_{23} \\
\begin{tikzcd}\scalebox{1.5}{$\circlearrowleft$}\end{tikzcd}\text{EP-}\texthelv{iii}&\iff& \mathrm{\bf M}_{14}\\
\begin{tikzcd}\scalebox{1.5}{$\circlearrowleft$}\end{tikzcd}\text{EP-}\texthelv{iv}&\iff& \mathrm{\bf M}_{34}
 \end{matrix}    
 \ .
\end{align}
We note that the red-dashed contour that encircles both EP-\texthelv{ii} and -\texthelv{iii} in Fig.~\ref{fig_mc} is the one that encloses the trivial two independent EP2s case.

\subsection{Discussion}

Currently, high-order-EP based apparatuses are considered as a rising candidate for prospective ultra-high-resolution sensor applications. The critical point of these applications is how exacting higher-order EPs can be achieved in experimental setups, guaranteed by symmetries. If the symmetries are not perfect, higher-order EPs split into several lower-order EPs or EP2s such as in the effective Hamiltonian of Eq.~(\ref{pertH}) with non-zero disorder strength. In the case of multiple EP2s, a high sensitivity, which is expected from a higher-order EP, cannot be achieved when perturbation $\delta$ is very small. We can separate the sensitivities from a higher-order EP and multiple EP2s via the energy or frequency differences following the sensitivity (see Supplementary Information).

Recently, a growing number of exceptional points have been realized in diverse research fields such as quantum many-body spectra, photonic crystals, optical waveguides, and non-Hermitian lattice models. This growth in observations of a large number of exceptional points is natural because of the ubiquity of 2nd order exceptional points in two-dimensional parameter spaces. In this work, we have shown that the topological structures around an arbitrary number of arbitrary-order EPs are permutation groups, and we have also systematically studied all possible perturbation-induced topological structures as classes in the groups. Finally, we confirmed our results in a non-Hermitian effective Hamiltonian and desymmetrized optical microcavities. We expect our results to open up a new field of topological classifications associated with tangled non-Hermitian singularities in general non-Hermitian systems.

\section{Method}

\subsection{Multi-valued-parameterization of the fully asymmetric cavity boundary}

Our asymmetric optical microcavity consists of a dielectric material which is defined inside a non-circular boundary shape with no rotational or mirror-reflection symmetry, given as follows:

\begin{equation}\label{eq:system}
    \begin{aligned}
    \frac{r(\theta;\vec{\eta})}{R} &= 1+\sum_{\{j=\pm\}}\frac{\varepsilon^j}{4}\left[1 +  f^j_\triangle(\delta^j,\phi^j,\theta) f^j_\square(\delta^j,\phi^j,\theta)\right]\ ,\\
    f_\triangle^\pm &\equiv \pm \left( 1 - \frac{2}{\pi}\arccos\left[(\delta^\pm-1)\cos\left( \frac{\theta}{2}\pm\frac{\phi^\pm}{4}\right)\right] \right)\ ,\\
    f_\square^\pm  &\equiv \frac{2}{\pi}\arctan\left[\frac{\sin\left(\frac{\theta}{2}\pm\frac{\phi^\pm}{4}\right)}{\delta^\pm}\right]\ ,
    \end{aligned}
\end{equation}
where $\vec{\eta}=\{\varepsilon^\pm,\delta^\pm,\phi^\pm\}$ is the system parameter. The set of equations in \eq{eq:system} describes an averaged-two-notch spiral, in which the two notches face one another in the opposite direction. To control the system morphology by only one independent parameter $\varepsilon^+$, we preliminarily fix five of the six total parameters as $\varepsilon^-=0.4\varepsilon^+$, $\delta^-=\delta^+=0.05$, and $\phi^\pm=\pi/20$. With this shape parameter, we adjust the material parameter so that our parameter space is spanned by $(n,\varepsilon\equiv\varepsilon^+)$. Note that $f^\pm_\triangle$ and $f^\pm_\square$ are periodic-triangle and -square functions, respectively. In Fig.~\ref{fig_mc}\tbf{a}, three different cases of the boundary shapes are illustrated: (1) rotational-symmetric when $\varepsilon=0$, (2) mirror-symmetric when $\varepsilon^+=\varepsilon^-\ne 0,\ \delta^+=\delta^-$, and (3) asymmetric when $\varepsilon^+\ne\varepsilon^-\ne 0,\ \delta^+\ne\delta^-$. Note that $\phi^\pm$ is only concerned with the relative position of the two notches.

\subsection{Optical modes}

Given the boundary shape of \eq{eq:system}, we obtain the optical modes by solving the two-dimensional Maxwell equations reduced to the Helmholtz wave equation,

\begin{align}
-\nabla^2\psi=n^2(\mathbf{r})k^2\psi\ ,
\label{eq:helm}
\end{align}
imposing a transverse-magnetic [TM; $\psi=(0,0,E_z)$] dielectric boundary condition at the cavity--vacuum interface. Here, $n(\mathbf{r})$ is the piecewise constant refractive index, and $k=\omega/c$ is a vacuum wavenumber with a time-harmonic frequency $\omega$ of $e^{-i\omega t}$, the speed of light $c$, and $i^2=-1$. Our microdisk is made of a (non-absorbing and non-magnetic) dielectric material having a refractive index of $n\sim 1.63$. We focus on the transverse-magnetic $[TM;\ \psi=(0,0,E_z)]$ polarization of the modes.

\subsubsection{Analytical circular cavity modes}

 In a circular cavity with radius $R$, the modes can be computed analytically in polar coordinates $(r,\theta)$ as
\begin{align}
 \Psi(r,\theta;nk,\omega)=\psi_m(nkr)\exp{(im\theta)}e^{-i\omega t}\ ,
\end{align}
~\cite{cao-wiersig}, where $\psi_m(nkr)\sim J_m(nkr)$ for \{$n>1,\ r<R$\} and $\psi_m(nkr)\sim H^{(1)}_m(kr)$ for \{$n=1,\ r>R$\}. Here, $J_m$ and $H_m^{(1)}$ denote the Bessel and the first kind Hankel functions of order $m$. Along with the azimuthal modal index $m$, we can define a radial modal index $l$ as well by counting the anti-nodes of $J_m$ inside the cavity. Hence, the optical modes in a circular cavity can be labeled by the mode index $(l,m)$.

The isolated microcavity systems, which are embedded in an infinite-extended vacuum environment, are open systems. Accordingly, the optical modes in the system are temporal decaying states characterized by a complex-valued $k=\omega/c\in\mathbb{C}$, where $\text{Im}(\omega) <0\to \text{Im}(k)<0$. That is, our solutions $k$s are in the fourth quadrant on the complex plane. We use its dimensionless form $kR$ scaled by an arbitrary characteristic length of the system. In the present work, the radius $R$ of the non-deformed cavity is taken for this role (i.e., $r=R$ when $\varepsilon=0$ in our asymmetric deformed optical microcavity).

\subsubsection{Numerical method for deformed cavity modes}
	
    To numerically compute the optical modes in our two-dimensional asymmetric deformed microdisk, we implement the so-called boundary element method (BEM)~\cite{bem}.
	
	The essential idea of the BEM relies on combining the free-space Green function $G$ and the unknown solution $\psi$ of the Helmholtz equations as
	\begin{equation}
	\begin{aligned}
		\left[\nabla^2+n^2k^2\right]G\left(r,r';k\right)&=\delta(r-r')\ ,\\ \left[\nabla^2+n^2k^2\right]\psi\left(r;k\right)&=0\ ,	    
    \end{aligned}
	\end{equation}
	for a domain enclosed by a simple connected closed envelope $\Gamma$. Here, $(r,r')$ is inside $\Gamma$, $G(r,r';k)=-i/4 H_0^{(1)}(nk|r-r'|)$ in two dimensions, and $\delta(r-r')$ is the two-dimensional Dirac delta function, where $H_0^{(1)}$ is the zeroth-order first kind Hankel function. The two equations are nested to derive a boundary integral equation (BIE) for unknown $\psi(s)$, $\partial_\nu\psi(s)$, and $k$, as follows: 
	\begin{align}
		\psi(r')=\oint_\Gamma \deri s\left[\psi(s)\partial_\nu G\left(s,r';k\right)-G\left(s,r';k\right)\partial_\nu\psi(s) \right]\nonumber.
	\end{align}
	Here, $r'$ is an arbitrary point inside $\Gamma$ with $s \subset \Gamma$, and $\partial_\nu$ denotes the outward normal derivative. By pushing $r'$ to be involved in $\Gamma$ too, as $r'\to s'\subset \Gamma$, we can complete the BIE formulation. The singular point of $G(s,s'=s)$ can be detoured in the sense of the Cauchy principal value.
	
	By meshing $\Gamma$ into $N$ number of segments, the BIE turns into a $2N \times 2N$ matrix equation of the BEM, 
	\begin{align}
		M\left(k\right)\Psi=0\ .
		\label{eq:BEM}
	\end{align}
	To solve this equation for $k$ and corresponding vectors $\Psi$, we take an iterative solution hunting strategy proposed, e.g., in~\cite{veble}.

\subsubsection{Tracing multiple eigenstates encircling EPs}
    It is found that numerically distinguishing the four solutions $\{k_1,k_2,k_3,k_4\}$ in the Riemann surface is a very challenging task, as their pairwise differences $|k_i-k_j|<\Delta,\ 1\le i,j \le 4$ are very small as $\Delta < 10^{-4}$. To work around this issue, we compute the four solutions simultaneously in terms of multiple column vectors, $u_0$, by applying a pseudo-Krylov orthonormalization~\cite{krylov} accompanied by a biorthogonality of optical modes. It is emphasized that although this process gives self-consistent, near-exact solutions that are well separated from one another, we have to perform one more refinement iteration to make these values converge to the numerically exact ones. The reason for this is because the normalization of modes in an open system is basically ill-conditioned, associated with phase rigidity~\cite{rigid}; this remains an unresolved issue, particularly when the modes are in the vicinity of EPs.
    
\section{Data availability}
The data that support the findings of this study are available from the corresponding authors upon reasonable request.

\section{Code availability}
The numerical codes used in this paper are available from the corresponding authors
upon reasonable request.

\section{acknowledgments}
We acknowledge financial support from the Institute for Basic Science in the Republic of Korea through the project IBS-R024-D1.

\section{Author contributions}
J.-W.R., J.-H.H., and C.-H.Y. conceived the original idea and carried out the theoretical analysis. C.-H.Y. performed the numerical analyses of the deformed microcavity. All authors discussed the manuscript and contributed to the writing.

\section{Competing interests}
The authors declare no competing interests.


\clearpage
	\newpage

\begin{thebibliography}{150}
\bibitem{Moi11} Moiseyev, N. {\it Non-Hermitian Quantum Mechanics} (Cambridge University Press, Cambridge, 2011).

\bibitem{Gan18} El-Ganainy, R., Makris, K. G., Khajavikhan, M., Musslimani, Z. H., Rotter, S. \& Christodoulides, D. N. Non-Hermitian physics and PT symmetry. {\it Nat. Phys.} {\bf 14}, 11-19 (2018).

\bibitem{Ash20} Ashida, Y., Gong, Z. \& Ueda, M. Non-Hermitian Physics. {\it Advances in Physics} {\bf 69}, 249-435 (2020).

\bibitem{Ber21} Bergholtz, E. J., Budich, J. C. \& Kunst, F. K. Exceptional Topology of Non-Hermitian Systems. {\it Rev. Mod. Phys.} {\bf 93}, 015005 (2021).

\bibitem{Kat66} Kato, T. {\it Perturbation theory of linear operators} (Springer, Berlin, 1966).

\bibitem{Hei90} Heiss, W. D. \& Sannino, A. L. Avoided level crossing and exceptional points. {\it J. Phys. A} {\bf 23}, 1167–1178 (1990).

\bibitem{Hei04} Heiss, W. D. Exceptional points of non-Hermitian operators. {\it J. Phys. A} {\bf 37}, 2455–2464 (2004).

\bibitem{Lee08} Lee, S.-Y., Ryu, J.-W., Shim, J.-B., Lee, S.-B., Kim, S. W. \& An, K. Divergent Petermann factor of interacting resonances in a stadium-shaped microcavity. {\it Phys. Rev. A} {\bf 78}, 015805 (2008).

\bibitem{Dem01} Dembowski, C., Gräf, H.-D., Harney, H. L., Heine, A., Heiss, W. D., Rehfeld, H., \& Richter, A. Experimental Observation of the Topological Structure of Exceptional Points. {\it Phys. Rev. Lett.} {\bf 86}, 787 (2001).

\bibitem{Mia16} Miao, P., Zhang, Z., Sun, J., Walasik, W., Longhi, S., Litchinitser, N. M. \& Feng, L. Orbital angular momemtum microlaser. {\it Science}, {\bf 353}, 464-467 (2016).

\bibitem{Yan18} Yang, H., Wang, C., Yu, T., Cao, Y. \& Yan, P. Antiferromagnetism Emerging in a Ferromagnet with Gain. {\it Phys. Rev. Lett.} {\bf 121}, 197201 (2018).

\bibitem{Yan21} Yang, Z., Schnyder, A. P., Hu, J. \& Chiu, C.-K. Fermion Doubling Theorems in Two-Dimensional Non-Hermitian Systems for Fermi Points and Exceptional Points. {\it Phys. Rev. Lett.} {\bf 126}, 086401 (2021).

\bibitem{Mir19} Miri, M.-A. \& Alù, A. Exceptional points in optics and photonics. {\it Science} {\bf 363}, eaar7709 (2019).

\bibitem{Che17} Chen, W., {\"O}zdemir, Ş. K., Zhao, G., Wiersig, J. \& Yang, L. Exceptional points enhance sensing in an optical microcavity. {\it Nature}, {\bf 548}, 192–196 (2017).

\bibitem{Car09} Cartarius, H., Main, J. \& Wunner, G. Exceptional points in the spectra of atoms in external fields. {\it Phys. Rev. A} {\bf 79}, 053408 (2009).

\bibitem{Ryu12} Ryu, J.-W., Lee, S.-Y. \& Kim, S. W. Analysis of multiple exceptional points related to three interacting eigenmodes in a non-Hermitian Hamiltonian. {\it Phys. Rev. A} {\bf 85}, 042101 (2012).

\bibitem{Zho18} Zhong, Q., Khajavikhan, M., Christodoulides, D. N. \& El-Ganainy, R. Winding around non-Hermitian singularities. {\it Nat. Commun.} {\bf 9}, 4808 (2018).

\bibitem{Din15} Ding, K., Zhang, Z. Q. \& Chan, C. T. Coalescence of exceptional points and phase diagrams for one-dimensional PT-symmetric photonic crystals. {\it Phys. Rev. B} {\bf 92}, 235310 (2015).

\bibitem{Din16} Ding, K., Ma, G., Xiao, M., Zhang, Z. Q. \& Chan, C. T. Emergence, Coalescence, and Topological Properties of Multiple Exceptional Points and Their Experimental Realization. {\it Phys. Rev. X} {\bf 6}, 021007 (2016).

\bibitem{Cui19} Cui, X., Ding, K., Dong, J.-W. \& Chan, C. T. Exceptional points and their coalescence of PT-symmetric interface states in photonic crystals. {\it Phys. Rev. B} {\bf 100}, 115412 (2019).

\bibitem{Jia20} Jia, J., Zhu, B., Ye, F., Zhong, H. \& Deng, H. Floquet higher-order exceptional points and dynamics in PT-symmetric quadrimer waveguides. {\it Eur. Phys. J. D} {\bf 74}, 117 (2020).

\bibitem{Tan20} Tang, W., Jiang, X., Ding, K., Xiao, Y.-X., Zhang,Z.-Q. Chan, C. T. \& Ma, G. Exceptional nexus with a hybrid topological invariant. {\it Science} {\bf 370}, 1077–1080 (2020).

\bibitem{Lui19} Luitz, D. \& Piazza, F. Exceptional points and the topology of quantum many-body spectra. {\it Phys. Rev. Res.} {\bf 1}, 033051 (2019).

\bibitem{Hei08} Heiss, W. D. Chirality of wavefunctions for three coalescing levels. {\it J. Phys. A: Math. Theor.} {\bf 41}, 244010 (2008).

\bibitem{Dem12} Demange, G. \& Graefe, E.-M. Signatures of three coalescing eigenfunctions. {\it J. Phys. A: Math. Theor.} {\bf 45}, 025303 (2012).

\bibitem{Hei16} Heiss, W. D. \& Wunner, G. A model of three coupled wave guides and third order exceptional points. {\it J. Phys. A: Math. Theor.} {\bf 49}, 495303 (2016).

\bibitem{Sch17} Schnabel, J., Cartarius, H., Main, J., Wunner, G. \& Heiss, W. D. PT-symmetric waveguide system with evidence of a third-order exceptional point. {\it Phys. Rev. A} {\bf 95}, 053868 (2017).

\bibitem{Hod17} Hodaei, H., Hassan, A. U., Wittek, S., Garcia-Gracia, H., El-Ganainy, R., Christodoulides, D. N. \& Khajavikhan, M. Enhanced sensitivity at higher-order exceptional points. {\it Nature} {\bf 548}, 187-191 (2017).

\bibitem{Nad17} Nada, M. Y., Othman, M. A. K. \& Capolino, F. Theory of coupled resonator optical waveguides exhibiting high-order exceptional points of degeneracy. {\it Phys. Rev. B} {\bf 96}, 184304 (2017). 

\bibitem{Kul18} Kullig, J., Yi, C.-H. \& Wiersig, J. Exceptional points by coupling of modes with different angular momenta in deformed microdisks: A perturbative analysis. {\it Phys. Rev. A} {\bf 98}, 023851 (2018).

\bibitem{Kul19} Kullig, J. \& Wiersig, J. High-order exceptional points of counterpropagating waves in weakly deformed microdisk cavities. {\it Phys. Rev. A} {\bf 100}, 043837 (2019).

\bibitem{Xia19} Xiao, Z., Li, H., Kottos, T. \& Alù, A. Enhanced Sensing and Nondegraded Thermal Noise Performance Based on PT-Symmetric Electronic Circuits with a Sixth-Order Exceptional Point. {\it Phys. Rev. Lett.} {\bf 123}, 213901 (2019).

\bibitem{Wan19} Wang, S., Hou, B., Lu, W., Chen, Y., Zhang, Z. Q. \& Chan, C. T. Arbitrary order exceptional point induced by photonic spin–orbit interaction in coupled resonators. {\it Nat. Commun.} {\bf 10}, 832 (2019).

\bibitem{Zha20} Zhang, S. M., Zhang, X. Z., Jin, L. \& Song, Z. High-order exceptional points in supersymmetric arrays. {\it Phys. Rev. A} {\bf 101}, 033820 (2020). 

\bibitem{Xia20} Xiao, Y.-X., Ding, K., Zhang, R.-Y., Hang, Z. H. \& Chan, C. T. Exceptional points make an astroid in non-Hermitian Lieb lattice: Evolution and topological protection. {\it Phys. Rev. B} {\bf 102}, 245144 (2020). 

\bibitem{Yu20} Yu, T., Yang, H., Song, L., Yan, P. \& Cao, Y. Higher-order exceptional points in ferromagnetic trilayers. {\it Phys. Rev. B} {\bf 101}, 144414 (2020).
\bibitem{Qui19} Quiroz-Juárez, M. A., Perez-Leija, A., Tschernig, K., Rodríguez-Lara, B. M., Magaña-Loaiza, O. S., Busch, K., Joglekar, Y. N. \& León-Montiel, R. J. Exceptional points of any order in a single, lossy waveguide beam splitter by photon-number-resolved detection. {\it Photon. Res.} {\bf 7}, 862-867 (2019).

\bibitem{Zho20} Zhong, Q., Kou, J., Özdemir, Ş. K. \& El-Ganainy, R. Hierarchical Construction of Higher-Order Exceptional Points. {\it Phys. Rev. Lett.} {\bf 125}, 203602 (2020).

\bibitem{Wan20} Wang, X.-Y., Wang, F.-F. \& Hu, X.-Y. Waveguide-induced coalescence of exceptional points. {\it Phys. Rev. A} {\bf 101}, 053820 (2020).

\bibitem{Alv18} Martinez Alvarez, V. M., Barrios Vargas, J. E. \& Foa Torres, L. E. F. Non-Hermitian robust edge states in one dimension: Anomalous localization and eigenspace condensation at exceptional points. {\it Phys. Rev. B} {\bf 97}, 121401 (2018).

\bibitem{Ham62} Hamermesh, M. {\it Group Theory and Its Application to Physical Problems} (Addison-Wesley Publishing Company, 1962).

\bibitem{wdheiss} Heiss, W. D. Chirality of wavefunctions for three coalescing levels. {\it J. Phys. A: Math. Theor.} {\bf 41}, 244010 (2008).

\bibitem{Ben98} Bender, C. M. \& Boettcher, S. Real Spectra in Non-Hermitian Hamiltonians Having PT Symmetry. {\it Phys. Rev. Lett.} {\bf 80}, 5243 (1998).

\bibitem{oeis} This is the same as the number of conjugacy classes in permutation group $S_n$ for some $n$. See https://oeis.org/A000041.

\bibitem{sgol} Gwak, S., Kim, H., Yu, H.-H., Ryu, J., Kim, C.-M. \& Yi, C.-H. Rayleigh scatterer-induced steady exceptional points of stable-island modes in a deformed optical microdisk. {\it Opt. Lett.} {\bf 46}, 2980-2983 (2021).

\bibitem{ep-prl} Yi, C.-H., Kullig, J. \& Wiersig, J. Pair of Exceptional Points in a Microdisk Cavity under an Extremely Weak Deformation. {\it Phys. Rev. Lett.} {\bf 120}, 093902 (2018).

\bibitem{fermi} Yi, C.-H., Yu, H.-H., Lee, J.-W. \& Kim, C.-M. Fermi resonance in optical microcavities. {\it Phys. Rev. E} {\bf 91}, 042903 (2015).

\bibitem{goos} Schomerus, H. \& Hentschel, M. Correcting ray optics at curved dielectric microresonator interfaces: Phase-space unification of Fresnel filtering and the Goos-H\"{a}nchen shift. {\it Phys. Rev. Lett.} {\bf 96}, 243903 (2006).

\bibitem{bem} Wiersig, J. Boundary element method for resonances in dielectric microcavities. {\it J. Opt. A: Pure Appl. Opt.} {\bf 5}, 53 (2002).

\bibitem{cao-wiersig} Cao, H., \& Wiersig, J. Dielectric microcavities: Model systems for wave chaos and non-Hermitian physics. {\it Rev. Mod. Phys.} {\bf 87}, 61-111 (2015).

\bibitem{veble} Veble, G., Prosen, T. \& Robnik, M. Expanded boundary integral method and chaotic time-reversal doublets in quantum billiards. {\it New J. Phys.} {\bf 9}, 15 (2007).

\bibitem{krylov} Meurant, G. \& Tebbens, J. D. Krylov methods for nonsymmetric linear systems - From theory to computations. {\it Springer Series in Computational Mathematics} {\bf 57} (2020).

\bibitem{rigid} Rotter, I. A non-Hermitian Hamilton operator and the physics of open quantum systems. {\it J. Phys. A: Math. Theor.} {\bf 42} 153001 (2009).
\end{thebibliography}
\end{document}


\draft
\title{Supplementary Information for Classification of multiple arbitrary-order non-Hermitian singularities}
\author{Jung-Wan Ryu, Jae-Ho Han, and Chang-Hwan Yi}
\date{\today}

\maketitle

\tableofcontents

\section{Signed holonomy matrices}

Describing the evolution of eigenstates with geometric phases after parametric variation along a loop encircling a 2nd order exceptional point (EP) once, the signed holonomy matrix connecting two eigenstates can be written as
\begin{eqnarray}
{\mathrm{\bf M}_{12}^{\pm}} = 
\begin{pmatrix} 
 0 & \pm 1 \\
 \mp 1 & 0 \\
 \end{pmatrix}.
\end{eqnarray}
Contrary to unsigned holonomy matrices, signed holonomy matrices satisfy the following relations,
\begin{eqnarray}
    (\mathrm{\bf M}_{ij}^{\pm})^2 = - \mathrm{\bf I}, \\
    \mathrm{\bf M}_{ij}^{\pm} = \mathrm{\bf M}_{ji}^{\mp} \neq \mathrm{\bf M}_{ji}^{\pm},
\end{eqnarray}
where $I$ is an identity matrix. As a result, the number of elements of the set for the eigenvectors is double that for the eigenvalues. A set for the eigenvalues is \{1,2\} in a two-state Hamiltonian but a set for the eigenvectors is \{-2,-1,1,2\} with a signed holonomy matrix as follows,
\begin{equation}
\begin{pmatrix} 
 -2 & -1 & 1 & 2 \\
 -1 & 2 & -2 & 1 \\
\end{pmatrix}
~~~\mathrm{for}~~~
\mathrm{\bf M}^{+}_{12}=
\begin{pmatrix} 
 0 & 1 \\
 -1 & 0 \\
\end{pmatrix},
\end{equation}
where the four elements \{-2,-1,1,2\} correspond to
\begin{equation}
    \left\{
    \begin{pmatrix} 
    0\\
    -1\\
    \end{pmatrix},
    \begin{pmatrix} 
    -1\\
    0\\
    \end{pmatrix},
    \begin{pmatrix} 
    1\\
    0\\
    \end{pmatrix},
    \begin{pmatrix} 
    0\\
    1\\
    \end{pmatrix}
    \right\}.
\end{equation}
The minus sign represents the geometrical phase gained during the encircling of the EPs.

The signed holonomy matrix connecting the four eigenstates can be written as
\begin{eqnarray}
{\mathrm{\bf M}_{12}^{\pm}} = 
\begin{pmatrix} 
 0 & \pm 1 & 0 & 0 \\
 \mp 1 & 0 & 0 & 0 \\
 0 & 0 & 1 & 0 \\
 0 & 0 & 0 & 1 \\
 \end{pmatrix}, \ \ \
 {\mathrm{\bf M}_{23}^{\pm}} = 
\begin{pmatrix} 
 1 & 0 & 0 & 0 \\
 0 & 0 & \pm 1 & 0 \\
 0 & \mp 1 & 0 & 0 \\
 0 & 0 & 0 & 1 \\
 \end{pmatrix},\ \ \
 {\mathrm{\bf M}_{34}^{\pm}} = 
\begin{pmatrix} 
 1 & 0 & 0 & 0 \\
 0 & 1 & 0 & 0 \\
 0 & 0 & 0 & \pm 1 \\
 0 & 0 & \mp 1 & 0 \\
 \end{pmatrix}, \\
 {\mathrm{\bf M}_{14}^{\pm}} = 
\begin{pmatrix} 
 0 & 0 & 0 & \pm 1 \\
 0 & 1 & 0 & 0 \\
 0 & 0 & 1 & 0 \\
 \mp 1 & 0 & 0 & 0 \\
 \end{pmatrix},\ \ \
 {\mathrm{\bf M}_{13}^{\pm}} = 
\begin{pmatrix} 
 0 & 0 & \pm 1 & 0 \\
 0 & 1 & 0 & 0 \\
 \mp 1 & 0 & 0 & 0 \\
 0 & 0 & 0 & 1 \\
 \end{pmatrix}, \ \ \
  {\mathrm{\bf M}_{24}^{\pm}} = 
\begin{pmatrix} 
 1 & 0 & 0 & 0 \\
 0 & 0 & 0 & \pm 1 \\
 0 & 0 & 1 & 0 \\
 0 & \mp 1 & 0 & 0 \\
 \end{pmatrix}. \nonumber
\end{eqnarray}
\section{Three-state case}

We consider multiple 2nd order EPs corresponding to a 3rd order EP in a $3 \times 3$ effective Hamiltonian. Describing the evolution of the eigenstates after parametric variation along a loop encircling a 2nd order EP once, the holonomy matrix connecting three eigenstates can be written as
\begin{eqnarray}
{\mathrm{\bf M}_{12}} = 
\begin{pmatrix} 
 0 & 1 & 0 \\
 1 & 0 & 0 \\
 0 & 0 & 1 \\
 \end{pmatrix}, \ \ \
{\mathrm{\bf M}_{23}} = 
\begin{pmatrix} 
 1 & 0 & 0 \\
 0 & 0 & 1 \\
 0 & 1 & 0 \\
 \end{pmatrix}, \ \ \
{\mathrm{\bf M}_{13}} = 
\begin{pmatrix} 
 0 & 0 & 1 \\
 0 & 1 & 0 \\
 1 & 0 & 0 \\
 \end{pmatrix}.
\end{eqnarray}
$\mathrm{\bf M}_{ij}$ describes that the $i$-th and $j$-th states are switched while the $k$-th state does not change after encircling $\mathrm{EP}_{ij}$. First, we consider the one EP case. The initial set $\{1,2,3\}$ should change into one of three possible outcome sets, $\{2,1,3\}$, $\{1,3,2\}$, or $\{3,2,1\}$, which are the outcomes of the three holonomy matrices $M_{12}$, $M_{23}$, and $M_{13}$, respectively. The three holonomy matrices have the same eigenvalues, ($-$1,1,1), corresponding to the same secular equation, $(\lambda^2 -1)(\lambda -1) = 0$. The result means that the three possible outcome sets are the same class, of which two states related to the EP exchange their eigenvalues after encircling the EP. When one more EP is added, the two EPs generate a cycle of length 3 for encircling two EPs. The six possible outcome sets are the same class. Adding one more EP, six possible outcome sets are the same class, of which two states exchange each other and one state returns itself after encircling the three EPs. Only two EPs and three EPs have corresponding 3rd order EPs. The classes, secular equations, eigenvalues, and the number of permutations according to the number of EPs are shown in Table~\ref{table_3}. The 15 possible permutations of the product of transpositions converge to only two classes, and then the secular equations and corresponding eigenvalues of the product of the holonomy matrices are
\begin{subequations}
\begin{align}
\label{sols3}
    (3^1) & : (\lambda^3 - 1) = 0, ~~~~~~~~~~~~ ((-1)^{2/3},-(-1)^{1/3},1), \\
    (1^1 2^1) & : (\lambda^2 - 1)(\lambda - 1) = 0, ~~~ (-1,1,1).
\end{align}
\end{subequations}

\begin{table*}
\caption{Number of 2nd order EPs (NEP), classes, corresponding secular equations, eigenvalues of the product of the holonomy matrices, and the number of possible permutations (NOP) for the three-state case. The asterisk represents that multiple EPs do not correspond to the 3rd order EP.}
\renewcommand{\arraystretch}{1.5}
\begin{center}
\scalebox{1.0}{%
\begin{tabular}{ |c|l|l|c|r| } 
\toprule
 NEP & Class & Secular Equation & Eigenvalues & NOP \\
\toprule
1* & $(1^1 2^1)$ & $(\lambda^2 -1)(\lambda -1) = 0$ & $(-1,1,1)$ & $3$ \\
\hline
2 & $(3^1)$ & $(\lambda^3 - 1) = 0$ & $((-1)^{2/3},-(-1)^{1/3},1)$ & $6$ \\
\hline
3 & $(1^1 2^1)$ & $(\lambda^2 -1)(\lambda -1) = 0$ & $(-1,1,1)$ & $6$ \\
\hline
\end{tabular}}
\end{center}
\label{table_3}
\end{table*}

\section{Notation of permutations for \texorpdfstring{$4\times4$}{4X4} effective Hamiltonian}

Holonomy matrices can be described by permutations of the set \{1,2,3,4\} of four states as
\begin{eqnarray}
{\mathrm{\bf M}_{12}} = 
\begin{pmatrix} 
 1 & 2 & 3 & 4 \\
 2 & 1 & 3 & 4 \\
\end{pmatrix} = 
( 1 2 ), \ \ \ 
 {\mathrm{\bf M}_{23}} = 
\begin{pmatrix} 
 1 & 2 & 3 & 4 \\
 1 & 3 & 2 & 4 \\
\end{pmatrix} = 
( 2 3 ), \ \ \ 
 {\mathrm{\bf M}_{34}} = 
\begin{pmatrix} 
 1 & 2 & 3 & 4 \\
 1 & 2 & 4 & 3 \\
\end{pmatrix} = 
( 3 4 ), \\\nonumber
 {\mathrm{\bf M}_{14}} = 
\begin{pmatrix} 
 1 & 2 & 3 & 4 \\
 4 & 2 & 3 & 1 \\
\end{pmatrix} = 
( 1 4 ), \ \ \
 {\mathrm{\bf M}_{13}} = 
\begin{pmatrix} 
 1 & 2 & 3 & 4 \\
 3 & 2 & 1 & 4 \\
\end{pmatrix} = 
( 1 3 ), \ \ \ 
  {\mathrm{\bf M}_{24}} = 
\begin{pmatrix} 
 1 & 2 & 3 & 4 \\
 1 & 4 & 3 & 2 \\
\end{pmatrix} = 
( 2 4 ),
\end{eqnarray}
in two-line notation and cyclic notation, respectively.

\section{Five-state case}

\begin{table*}
\caption{Number of 2nd order EPs (NEP), classes, corresponding secular equations, eigenvalues of the product of holonomy matrices, and the number of possible permutations (NOP) for the five-state case.}
\renewcommand{\arraystretch}{1.5}
\begin{center}
\scalebox{1.0}{%
\begin{tabular}{ |c|l|l|c|r| } 
\toprule
 NEP & Class & Secular Equation & Eigenvalues & NOP \\
\toprule
1 & $(1^3 2^1)$ & $(\lambda^2 - 1)(\lambda - 1)^3 = 0$ & $(-1,1,1,1,1)$ & $10$ \\
\hline
2 & $(1^2 3^1)$ & $(\lambda^3 - 1)(\lambda - 1)^2 = 0$ & $(1,1,1,-(-1)^{1/3},(-1)^{2/3})$ & $60$ \\
\hline
2 & $(1^1 2^2)$ & $(\lambda^2 - 1)^{2} (\lambda - 1) = 0$ & $(-1,-1,1,1,1)$ & $30$ \\
\hline
3 & $(1^3 2^1)$ & $(\lambda^2 - 1)(\lambda - 1)^3 = 0$ & $(-1,1,1,1,1)$ & $60$ \\
\hline
3 & $(1^1 4^1)$ & $(\lambda^4 - 1) (\lambda - 1) = 0$ & $(-1,i,-i,1,1)$ & $480$ \\
\hline
3 & $(2^1 3^1)$ & $(\lambda^3 - 1)(\lambda^2 - 1) = 0$ & $(-1,1,1,-(-1)^{1/3},(-1)^{2/3})$ & $180$ \\
\hline
4 & $(1^2 3^1)$ & $(\lambda^3 - 1)(\lambda - 1)^2 = 0$ & $(1,1,1,-(-1)^{1/3},(-1)^{2/3})$ & $1080$ \\
\hline
4 & $(1^1 2^2)$ & $(\lambda^2 - 1)^{2} (\lambda - 1) = 0$ & $(-1,-1,1,1,1)$ & $960$ \\
\hline
4 & $(5^1)$ & $(\lambda^5 - 1) = 0$ & $((-1)^{4/5},(-1)^{2/5},-(-1)^{1/5},-(-1)^{3/5},1)$ & $3000$ \\
\hline
5 & $(1^3 2^1)$ & $(\lambda^2 - 1)(\lambda - 1)^3 = 0$ & $(-1,1,1,1,1)$ & $1800$ \\
\hline
5 & $(1^1 4^1)$ & $(\lambda^4 - 1) (\lambda - 1) = 0$ & $(-1,i,-i,1,1)$ & $13320$ \\
\hline
5 & $(2^1 3^1)$ & $(\lambda^3 - 1)(\lambda^2 - 1) = 0$ & $(-1,1,1,-(-1)^{1/3},(-1)^{2/3})$  & $15120$ \\
\hline
6 & $(1^2 3^1)$ & $(\lambda^3 - 1)(\lambda - 1)^2 = 0$ & $(1,1,1,-(-1)^{1/3},(-1)^{2/3})$ & $34560$ \\
\hline
6 & $(1^1 2^2)$ & $(\lambda^2 - 1)^{2} (\lambda - 1) = 0$ & $(-1,-1,1,1,1)$ & $44160$ \\
\hline
6 & $(5^1)$ & $(\lambda^5 - 1) = 0$ & $((-1)^{4/5},(-1)^{2/5},-(-1)^{1/5},-(-1)^{3/5},1)$ & $72000$ \\
\hline
6 & $(1^5)$ & $(\lambda - 1)^{5} = 0$ & $(1,1,1,1,1)$ & $480$ \\
\hline
7 & $(1^3 2^1)$ & $(\lambda^2 - 1)(\lambda - 1)^3 = 0$ & $(-1,1,1,1,1)$ & $89040$ \\
\hline
7 & $(1^1 4^1)$ & $(\lambda^4 - 1) (\lambda - 1) = 0$ & $(-1,i,-i,1,1)$ & $297360$ \\
\hline
7 & $(2^1 3^1)$ & $(\lambda^3 - 1)(\lambda^2 - 1) = 0$ & $(-1,1,1,-(-1)^{1/3},(-1)^{2/3})$ & $218400$ \\
\hline
8 & $(1^2 3^1)$ & $(\lambda^3 - 1)(\lambda - 1)^2 = 0$ & $(1,1,1,-(-1)^{1/3},(-1)^{2/3})$ & $624720$ \\
\hline
8 & $(1^1 2^2)$ & $(\lambda^2 - 1)^{2} (\lambda - 1) = 0$ & $(-1,-1,1,1,1)$ & $447360$ \\
\hline
8 & $(5^1)$ & $(\lambda^5 - 1) = 0$ & $((-1)^{4/5},(-1)^{2/5},-(-1)^{1/5},-(-1)^{3/5},1)$ & $702240$ \\
\hline
8 & $(1^5)$ & $(\lambda - 1)^{5} = 0$ & $(1,1,1,1,1)$ & $40080$ \\
\hline
9 & $(1^3 2^1)$ & $(\lambda^2 - 1)(\lambda - 1)^3 = 0$ & $(-1,1,1,1,1)$ & $704880$ \\
\hline
9 & $(1^1 4^1)$ & $(\lambda^4 - 1) (\lambda - 1) = 0$ & $(-1,i,-i,1,1)$ & $1802880$ \\
\hline
9 & $(2^1 3^1)$ & $(\lambda^3 - 1)(\lambda^2 - 1) = 0$ & $(-1,1,1,-(-1)^{1/3},(-1)^{2/3})$ & $1121040$ \\
\hline
10 & $(1^2 3^1)$ & $(\lambda^3 - 1)(\lambda - 1)^2 = 0$ & $(1,1,1,-(-1)^{1/3},(-1)^{2/3})$ & $1299840$ \\
\hline
10 & $(1^1 2^2)$ & $(\lambda^2 - 1)^{2} (\lambda - 1) = 0$ & $(-1,-1,1,1,1)$ & $895680$ \\
\hline
10 & $(5^1)$ & $(\lambda^5 - 1) = 0$ & $((-1)^{4/5},(-1)^{2/5},-(-1)^{1/5},-(-1)^{3/5},1)$ & $1353600$ \\
\hline
10 & $(1^5)$ & $(\lambda - 1)^{5} = 0$ & $(1,1,1,1,1)$ & $79680$ \\
\hline
\end{tabular}}
\end{center}
\label{table_5}
\end{table*}

We consider multiple 2nd order EPs corresponding to a 5th order EP in a $5 \times 5$ effective Hamiltonian. Ten holonomy matrices can be defined for five states. Using the holonomy matrices, the classes, secular equations, eigenvalues, and the number of permutations according to the number of EPs for the five-state set are shown in Table~\ref{table_5}. Seven secular equations and corresponding eigenvalues are
\begin{subequations}
\begin{align}
    (5^1) &: (\lambda^5 - 1) = 0, ~~~~~~~~~~~~~~~ ((-1)^{\frac{4}{5}},(-1)^{\frac{2}{5}},-(-1)^{\frac{1}{5}},-(-1)^{\frac{3}{5}},1), \\
    (1^1 4^1) &: (\lambda^4 - 1)(\lambda -1) = 0, ~~~~~~ (-1,i,-i,1,1), \\
    (1^2 3^1) &: (\lambda^3 - 1)(\lambda - 1)^2 = 0, ~~~~~ (1,1,1,-(-1)^{\frac{1}{3}},(-1)^{\frac{2}{3}}),  \\
    (2^1 3^1) &: (\lambda^3 - 1)(\lambda^2 - 1) = 0, ~~~~~ (-1,1,1,-(-1)^{\frac{1}{3}},(-1)^{\frac{2}{3}}),  \\  
    (1^3 2^1) &: (\lambda^2 - 1)(\lambda-1)^3 = 0, ~~~~~ (-1,1,1,1,1),  \\
    (1^1 2^2) &: (\lambda^2 - 1)^2 (\lambda - 1) = 0, ~~~~~ (-1,-1,1,1,1), \\
    (1^5) &: (\lambda - 1)^5 = 0, ~~~~~~~~~~~~~~~~ (1,1,1,1,1).
\end{align}
\end{subequations}

\section{Equivalent loops depending on the choice of branch cuts and classes independent of the choice}

\begin{figure*}[b]
\begin{center}
\includegraphics[width=\figsizetwo\textwidth]{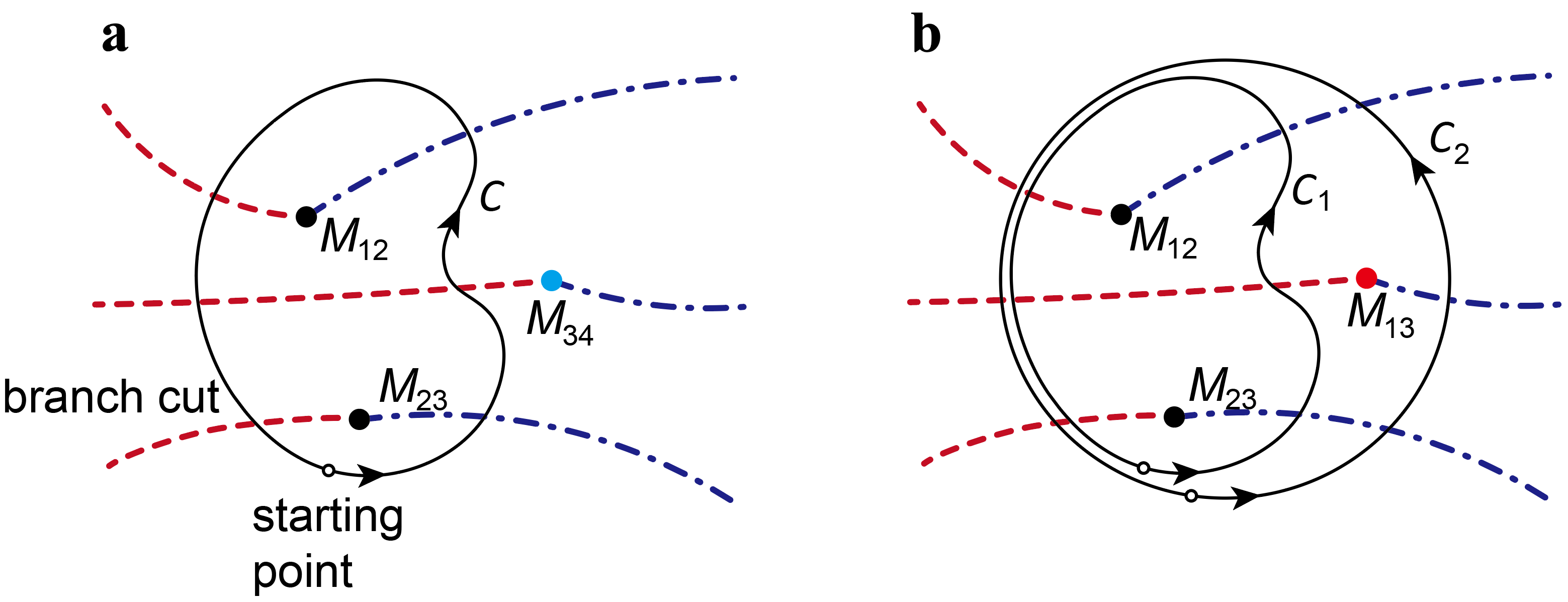}
\caption{
{\bf a.} Encircling two EPs (black dots) along the curve $\mathcal C$ starts from the point denoted by an open circle. The conjugacy class of the stroboscopic encircling is independent of the branch cuts, red dashed lines, or blue dot-dashed lines. The EP outside of the curve (blue dot) is irrelevant. {\bf b.} Encircling two or three EPs along the curve $\mathcal C_1$ or $\mathcal C_2$, respectively. The stroboscopic encircling along curve $\mathcal C_1$ depends on the branch cuts because the EP outside $\mathcal C_1$ (red dot) is relevant. Including this EP, i.e., following the curve $\mathcal C_2$, the stroboscopic encircling becomes independent of the branch cuts again.
}
\label{bc}
\end{center}
\end{figure*}

In this section, we show that the conjugacy class of stroboscopic encircling is independent of the choice of branch cuts of EPs with some examples. Consider a system having three EP2s characterized by holonomy matrices $M_{12}$, $M_{23}$, and $M_{34}$, as shown in Fig.~\ref{bc}{\bf a}. The branch cuts can be chosen freely, and we choose two sets of branch cuts, the red dashed and blue dot-dashed lines shown in the figure. For the red branch cut case, the stroboscopic encircling along the curve $\mathcal C$ is given by
\begin{eqnarray}
\mathcal C \text{(red)} : M_{23} M_{34} M_{12} M_{34} = (132),
\end{eqnarray}
in cyclic notation. For the blue branch cut case, the homology matrix is changed to 
\begin{eqnarray}
\mathcal C \text{(blue)} : M_{12} M_{23} = (123).
\end{eqnarray}
One can recognize that although the stroboscopic encirclings are different, the conjugacy classes are the same. This result holds when the EPs outside of the curve are irrelevant. Here, irrelevant means that the branch cut of the EP can be deformed to lie outside of the curve without touching the EPs. For example, for the blue EP having holonomy matrix $M_{34}$ in Fig. \ref{bc}{\bf a} (blue dot), the red branch cut can be deformed into the blue branch cut by clockwise rotation. It seems to pass through the EP with $M_{12}$, but the branch cut separates the third and fourth sheet while the EP is the coalescence point of the first and the second sheets, so they do not meet.

However, there can be an EP that is relevant. Consider the case of EPs and curves shown in Fig. \ref{bc}{\bf b}. The EP outside of curve $\mathcal C_1$ (red dot, characterized by $M_{13}$) is changed from the previous example. Along curve $\mathcal C_1$, the stroboscopic encirclings are
\begin{eqnarray}
&&\mathcal C_1 \text{(red)} : M_{23} M_{13} M_{12} M_{13} = (1), \\
&&\mathcal C_1 \text{(blue)} : M_{12} M_{23} = (123), \nonumber
\end{eqnarray}
for each branch cut. The conjugacy classes of the stroboscopic encircling are now different depending on the branch cuts. In this case, the red branch cut of the EP with $M_{13}$ (red dot in the figure) cannot pass through the EPs inside curve $\mathcal C_1$. If we encircle all three EPs, as with curve $\mathcal C_2$ in the figure, the conjugacy class of the stroboscopic encircling becomes independent of the choice of branch cuts. One can check this by calculating the stroboscopic encirclings,
\begin{eqnarray}
&&\mathcal C_2 \text{(red)} : M_{23} M_{13} M_{12} = (13), \\
&&\mathcal C_2 \text{(blue)} : M_{12} M_{13} M_{23} = (13), \nonumber
\end{eqnarray}
which are in the same conjugacy class (in this case, even the stroboscopic encircling themselves are the same).

The main text considers the stroboscopic encircling along the curve large enough to cover all the relevant EPs. Therefore, we can classify all the stroboscopic encirclings into conjugacy classes without concerning the branch cuts.  

\section{Cyclic building blocks for six and seven states}

Using cyclic building blocks, we can reassemble the classes of multiple EPs. We can symbolically classify the cases of six and seven states using the cyclic building blocks as follows:
\begin{itemize}
    \item 6 states: $\text{\textcircled{1}}\times6$, $\text{\textcircled{1}}\times4 + \text{\textcircled{2}}$, $\text{\textcircled{1}}\times3 + \text{\textcircled{3}}$, $\text{\textcircled{1}}\times2 + \text{\textcircled{2}}\times2$, $\text{\textcircled{1}}\times2 + \text{\textcircled{4}}$, $\text{\textcircled{1}} + \text{\textcircled{2}} + \text{\textcircled{3}}$, $\text{\textcircled{1}} + \text{\textcircled{5}}$, $\text{\textcircled{2}}\times2$, $\text{\textcircled{2}}$ + $\text{\textcircled{4}}$, $\text{\textcircled{3}}\times2$, and $\text{\textcircled{6}}$.
    \item 7 states: $\text{\textcircled{1}}\times7$, $\text{\textcircled{1}}\times5 + \text{\textcircled{2}}$, $\text{\textcircled{1}}\times4 + \text{\textcircled{3}}$, $\text{\textcircled{1}}\times3 + \text{\textcircled{4}}$, $\text{\textcircled{1}}\times3 + \text{\textcircled{2}}\times2$, $\text{\textcircled{1}}\times2 + \text{\textcircled{2}} + \text{\textcircled{3}}$, $\text{\textcircled{1}}\times2 + \text{\textcircled{5}}$, $\text{\textcircled{1}} + \text{\textcircled{2}}\times3$, $\text{\textcircled{1}} + \text{\textcircled{2}} + \text{\textcircled{4}}$, $\text{\textcircled{1}} + \text{\textcircled{3}}\times2$, $\text{\textcircled{1}} + \text{\textcircled{6}}$, $\text{\textcircled{2}}\times2 + \text{\textcircled{3}}$, $\text{\textcircled{2}} + \text{\textcircled{5}}$, $\text{\textcircled{3}} + \text{\textcircled{4}}$, and $\text{\textcircled{7}}$.
\end{itemize}

\section{Sensitivity of multiple 2nd order EPs}

\begin{figure*}[t]
\begin{center}
\includegraphics[width=\figsizeone\textwidth]{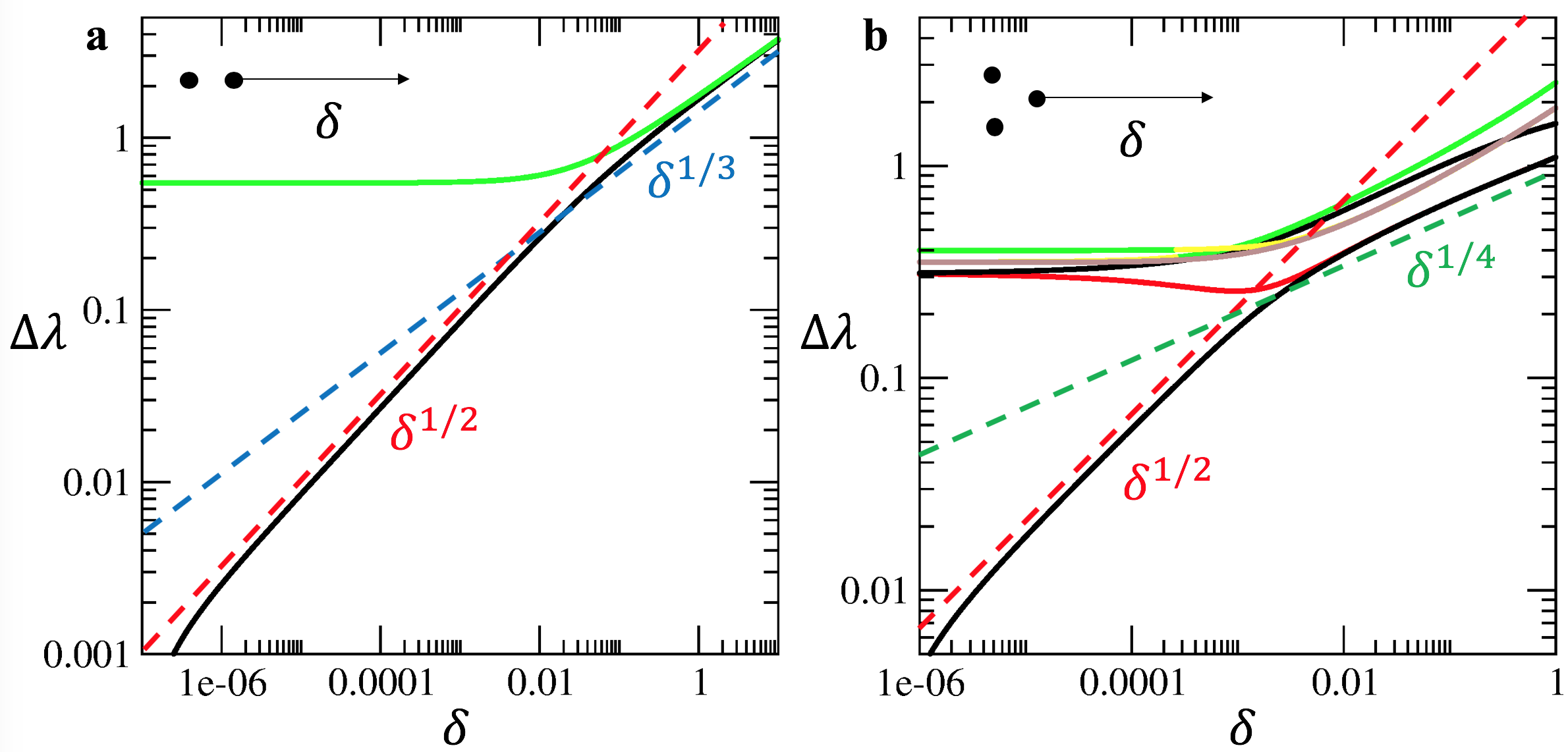}
\caption{{\bf a.} Sensitivity in a perturbed $3 \times 3$ effective Hamiltonian exhibiting two 2nd order EPs. {\bf b.} Sensitivity in a perturbed $4 \times 4$ effective Hamiltonian exhibiting three 2nd order EPs. Red, blue, and green dashed lines represent $\delta^{1/2}$, $\delta^{1/3}$, and $\delta^{1/4}$, respectively.
}
\label{sensitivity}
\end{center}
\end{figure*}

We add disorders to $3 \times 3$ and $4 \times 4$ perturbed Jordan normal forms 
\begin{eqnarray}
H &=& 
\begin{pmatrix} 
 0 & 1 & 0 \\
 d_3 & 0 & 1 \\
 \delta & 0 & 0 \\
 \end{pmatrix}
\end{eqnarray}
and
\begin{eqnarray}
H &=&
\begin{pmatrix} 
 0 & 1 & 0 & 0 \\
 \delta + d_4 & 0 & 1 & 0 \\
 \delta + d_4 & 0 & 0 & 1 \\
 \delta & 0 & 0 & 0 \\
 \end{pmatrix},
\end{eqnarray}
where $d_{3}=0.1$ and $d_{4}=0.01$, respectively. Adding the disorders, the 3rd and 4th order EPs are split into two and three 2nd order EPs in Fig.~\ref{sensitivity} {\bf a} and {\bf b}, respectively. Assuming one among multiple EPs as the origin, $\delta$ can be considered as a parameter perturbing the systems. In the $3 \times 3$ case, sensitivity against small perturbation changes from $\delta^{1/2}$ to $\delta^{1/3}$ as $\delta$ increases. While only two states are affected by the right EP in Fig.~\ref{sensitivity}{\bf a} when $\delta$ is very small, all states are affected by two EPs when $\delta$ is larger than $0.1$. In the $4 \times 4$ case, sensitivity against small perturbation changes from $\delta^{1/2}$ to $\delta^{1/4}$ as $\delta$ increases. While only two states are affected by the right EP in Fig.~\ref{sensitivity}{\bf b} when $\delta$ is very small, all states are affected by three EPs when $\delta$ is larger than $0.001$. As the splitting of multiple EPs approaches zero, i.e., in the case of higher-order EPs, the critical $\delta$ approaches zero and the regimes with $\delta^{1/2}$ disappear. As a result, there can be two possible reasons that we cannot achieve the $N$th order root of the perturbation strength in $N$th order EPs. First, the class of an $N$th order EP can have lower-order roots. Next, $N$th order EPs can have a lower-order root of the perturbation strength by imperfect implementation even though the class has the $N$th order root.  
